\documentclass[aps,pra,amsmath,amssymb,twocolumn,superscriptaddress,notitlepage]{revtex4-1}
\usepackage{graphicx}
\usepackage{color}
\usepackage{braket}
\usepackage{cases}
\definecolor{burntorange}{rgb}{0.8, 0.33, 0.0}\usepackage[colorlinks,citecolor=blue,linkcolor=red,urlcolor=blue]{hyperref}
\hypersetup{
  pdfkeywords={},
  pdftitle={Criticality-Enhanced Quantum Sensing via Continuous Measurement},
  }

\makeatletter

\newcommand{\ha}{\hat{c}}
\newcommand{\hadag}{\hat{c}^\dagger}

\newsavebox{\@brx}
\newcommand{\llangle}[1][]{\savebox{\@brx}{\(\m@th{#1\langle}\)}%
  \mathopen{\copy\@brx\kern-0.5\wd\@brx\usebox{\@brx}}}
\newcommand{\rrangle}[1][]{\savebox{\@brx}{\(\m@th{#1\rangle}\)}%
  \mathclose{\copy\@brx\kern-0.5\wd\@brx\usebox{\@brx}}}
\makeatother

\begin{document}

\title{Criticality-Enhanced Quantum Sensing via Continuous Measurement}

\author{Theodoros Ilias}
\thanks{T. I. and D. Y. contributed equally to this work.}
\author{Dayou Yang}
\email{dayou.yang@uni-ulm.de} 
\author{Susana F. Huelga} \email{susana.huelga@uni-ulm.de}
\author{Martin B. Plenio} \email{martin.plenio@uni-ulm.de}

\affiliation{
Institut f\"ur Theoretische Physik and IQST, Universit\"at Ulm,
Albert-Einstein-Allee 11, D-89069 Ulm, Germany}

\date{\today}

\begin{abstract}
Present protocols of criticality-enhanced sensing with open quantum sensors assume direct measurement of the sensor and omit the radiation quanta emitted to the environment, thereby potentially missing valuable information. Here we propose a protocol for criticality-enhanced sensing via continuous observation of the emitted radiation quanta. Under general assumptions, we establish a scaling theory for the global quantum Fisher information of the joint system and environment state at dissipative critical points. We derive universal scaling laws featuring transient and long-time behavior governed by the underlying critical exponents. Importantly, such scaling laws exceed the standard quantum limit and can in principle saturate the Heisenberg limit. To harness such advantageous scaling, we propose a practical sensing scheme based on continuous detection of the emitted quanta as realized experimentally in various quantum-optical setups. In such a scheme a single interrogation corresponds to a (stochastic) quantum trajectory of the open system evolving under the nonunitary dynamics dependent on the parameter to be sensed and the backaction of the continuous measurement. Remarkably, we demonstrate that the associated precision scaling significantly exceeds that based on direct measurement of the critical steady state, thereby establishing the metrological value of the continuous detection of the emitted quanta at dissipative criticality. We illustrate our protocol via counting the photons emitted by the open Rabi model, a paradigmatic model for the study of dissipative phase transition with finite components. Our protocol is applicable to generic quantum-optical open sensors permitting continuous readout, and may find applications at the frontier of quantum sensing, such as the human-machine interface, magnetic diagnosis of heart disease, and zero-field nuclear magnetic resonance.
\end{abstract}

\maketitle

\section{Introduction}
A most ambitious vista of quantum sensing is to enhance the sensor precision from the standard quantum limit (SQL)~\cite{Braginski__1975,PhysRevLett.45.75,RevModPhys.52.341} of independent, uncorrelated measurements towards the Heisenberg limit (HL)~\cite{PhysRevD.23.1693,Giovannetti1330}, the ultimate precision allowed by quantum mechanics. Given a fixed number $N$ of the sensor components and a total time $t$ of the sensing interrogation, the SQL represents a (classical) scaling of the estimation error as $\sim 1/\sqrt{N t}$, whereas the HL features a quantum-enhanced scaling $\sim 1/{N t}$~\cite{PhysRevA.46.R6797}. Such enhancement, on the one hand, can be achieved by preparing individual sensor components in entangled quantum states, e.g., the Greenberger--Horne--Zeilinger state~\cite{GHZ} and spin-squeezed states~\cite{PhysRevA.47.5138,PhysRevA.46.R6797}. Nevertheless, such states are technically challenging to scale, fragile when exposed to noise~\cite{PhysRevLett.79.3865}, and thus far restricted to a small number of components, see recent advances~\cite{Strobel424,Bohnet1297,Luo620,Omran570,Song574,kaubruegger2021quantum,marciniak2021optimal}. On the other hand, alternative strategies without direct preparation of entangled states open intriguing and promising routes towards quantum-enhanced sensing with affordable technological requirements; see Ref.~\cite{RevModPhys.90.035006} for an overview. 

A recent highlight among these is criticality-enhanced sensing~\cite{PhysRevA.78.042105,PhysRevA.88.021801,PhysRevX.8.021022,PhysRevA.96.013817,PhysRevLett.123.173601,PhysRevLett.124.120504,dicandia2021critical,Gietka2021adiabaticcritical,PhysRevLett.126.010502,Salado_Mej_a_2021}. The key ingredient there is the universally divergent susceptibility of the ground state to small parameter variation of the Hamiltonian at a quantum critical point (CP). Such divergence translates directly into a divergent quantum Fisher information (QFI). Protocols of criticality-enhanced sensing typically adopt one of the following two approaches. The first ~\cite{PhysRevA.88.021801,PhysRevLett.126.010502} is based on the time evolution of the ground state following a quench of the Hamiltonian parameters across the CP, i.e., exploiting the dynamic susceptibility. The precision is quantified by the QFI of the time-evolved state, which obeys a sub-Heisenberg scaling $\sim t^2 N^\alpha$ with $\alpha\leq 2$ determined by the critical exponents of the underlying CP. The second approach~\cite{PhysRevA.78.042105,PhysRevLett.124.120504} exploits the static susceptibility of the ground state, by adiabatically switching the Hamiltonian parameters across the CP. The precision is quantified by the QFI of the ground state, which may obey an apparent super-Heisenberg scaling with respect to $N$. This, however, comes at the price of a divergent interrogation time to maintain adiabaticity close to the CP~\cite{PhysRevX.8.021022}. Taking into account such time, the ground-state QFI obeys the same scaling $\sim t^2 N^\alpha$ as in the quench approach~\cite{PhysRevX.8.021022}. Therefore, the two approaches are equivalent and both provide a promising avenue towards the HL based on engineering short-range interacting many-body systems close to quantum criticality. Experimentally, such a capability has been demonstrated in various quantum-optical setups well isolated from the environment~\cite{RevModPhys.80.885, Blatt:2012aa, Houck:2012aa, Keesling:2019aa}. 

Open quantum-optical setups represent another interesting candidate for criticality-enhanced sensing. These systems support dissipative criticality~\cite{Baumann:2010wv,Klinder_2015,PhysRevLett.113.020408,PhysRevLett.118.247402,PhysRevX.7.011016,PhysRevX.7.011012}, defined via gap closing in the highest real part of the spectrum of the Liouville superoperator for the reduced density matrix of the system~\cite{Sieberer_2016}. At a dissipative CP, the thermodynamic properties of the steady density matrix manifest divergent scaling behavior, bearing similarities to the behavior of the ground state at a quantum CP~\cite{Sieberer_2016}. 

State-of-the-art protocols~\cite{PhysRevA.96.013817,PhysRevLett.123.173601,PhysRevLett.124.120504,dicandia2021critical} for critical open sensors quantify the sensor precision limit via the QFI of the reduced density matrix of the system. This, however, only represents the optimal precision achieved by direct measurement of the system.
In stark contrast to closed systems, open dissipative systems continuously exchange radiation quanta with their environments, which carry information about the system and, therefore, about the parameter to be sensed. Such radiation quanta may be detected continuously in time via mature experimental techniques in quantum optics, e.g., photon counting and homodyne measurement
~\cite{Purdy801,Hood1447,PhysRevLett.96.043003,Minev:2019aa}, accomplishing indirect monitoring of the system via measurement of the environment. Such a unique opportunity makes open sensors a natural platform for implementing continuous-measurement-based sensing schemes~\cite{PhysRevA.87.032115,PhysRevA.89.052110,PhysRevA.94.032103,PhysRevLett.112.170401,2015JPhA...48J5301C,PhysRevA.64.042105,PhysRevA.93.022103,Schmitt832,PhysRevA.101.032347} and, from a theoretical point of view, requires the use of the global QFI of the joint system and environment state $|\Psi(t)
\rangle$
\begin{equation}
\label{eq: gQFI_1}
I_\theta(t)=4[\langle\partial_\theta\Psi(t)|\partial_\theta\Psi(t)\rangle-|\langle\Psi(t)|\partial_\theta\Psi(t)\rangle|^2]
\end{equation}
as the ultimate precision bound, which can in principle be achieved by the most general measurement of the joint system and environment (implementing such measurement in practice, however, may be challenging). This has been emphasized in sensing with non-interacting open system~\cite{PhysRevA.87.032115,PhysRevA.89.052110,PhysRevA.94.032103,PhysRevLett.112.170401, 2015JPhA...48J5301C,PhysRevA.64.042105} and systems manifesting intermittent dynamics~\cite{PhysRevA.93.022103}, as well as in scenarios of noisy quantum sensing with observed environments~\cite{Albarelli2018restoringheisenberg,PhysRevA.93.032123,PhysRevLett.125.200505}. 
\begin{figure*}[t!]
\centering{} \includegraphics[width=0.98\textwidth]{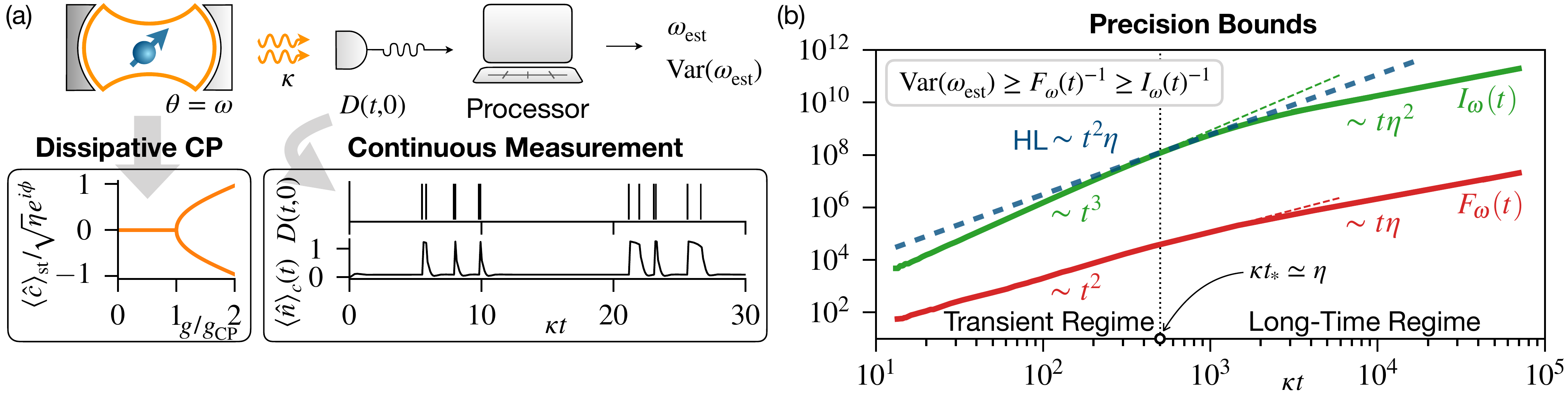} 
\caption{(a) The open Rabi model as a model system demonstrating criticality enhanced sensing via continuous measurement. We consider the sensing of the cavity mode frequency $\omega$ as an illustration. The cavity leaks photons at a rate $\kappa$, counted continuously by a photon detector. The model parameters are tuned to a dissipative CP $g_{\rm CP}$, where its steady state undergoes a continuous phase transition (see text). Here $\eta$ is the effective system size, $\phi=\arctan(2\kappa/\omega)$. A sensing interrogation corresponds to a quantum trajectory consisting of the continuously detected signal $D(t,0)$ and the associated conditional evolution of the open system, as exemplified by the cavity mode occupation $\langle \hat{n}\rangle_c(t)$. Processing $D(t,0)$ (e.g., via Bayesian inference~\cite{PhysRevA.87.032115,PhysRevA.87.032115,PhysRevA.89.052110,PhysRevA.94.032103}) provides an estimator $\omega_{\rm est}$ with an imprecision (variance) ${\rm Var}(\omega_{\rm est})$.
(b) Scaling of the precision bounds at dissipative criticality, including the global QFI of the joint system and environment $I_\omega(t)$, and the FI $F_\omega(t)$ of the detected signal. Both quantities manifest algebraic scaling with respect to $t$ and $\eta$, and for illustration we show the numerical results for $\eta=500$, $\omega=1$, $\kappa=0.1$. With respect to the interrogation time $t$, the precision bounds obeys (super-)Heisenberg scaling in the transient regime, $t\lesssim t_*\simeq\eta/\kappa$, whereas linear scaling in the long-time regime, $t\gg t_*$. The scaling should be contrasted with the HL $\sim t^2\eta$ (thick dashed blue) and the SQL $\sim t\sqrt{\eta}$ (not shown). The scaling exponents of the global QFI can be related to the critical exponents $z=1$, $\Delta_{\hat{n}}=-1/2$ and the spatial dimension $d=0$ via the general formulas \eqref{eq:QFI_short_time_general} and \eqref{eq:QFI_long_time_general}.}
\label{fig:fig1} 
\end{figure*}
In contrast to the existing studies~\cite{PhysRevA.87.032115,PhysRevA.89.052110,PhysRevA.94.032103,PhysRevLett.112.170401,2015JPhA...48J5301C,PhysRevA.64.042105,PhysRevA.93.022103,Schmitt832,PhysRevA.101.032347,Albarelli2018restoringheisenberg,PhysRevA.93.032123,PhysRevLett.125.200505}, here we are interested in open interacting many-body sensors at (continuous) dissipative CPs. Two fundamentally important questions emerge naturally in this context. (i) Does the global QFI \eqref{eq: gQFI_1} manifest criticality-enhanced universal scaling at a dissipative CP? (ii) If so, can we access such enhanced precision scaling via realistic measurement schemes of the emitted radiation quanta? In this paper, we provide positive answers to both questions. We show that the global QFI \eqref{eq: gQFI_1} for a generic Markovian open sensor obeys universal scaling laws at dissipative criticality, which are governed by and can be related analytically to the universal critical exponents of the underlying CP. Such scaling exceeds the SQL, and can in principle saturate the HL. Moreover, as a practical scheme to access the criticality-enhanced precision, we analyze a sensing protocol based on continuous measurements that is applicable to the the sensing of arbitrary parameters at generic dissipative CPs, relevant to diverse platforms across experimental quantum optics. 

Our key findings are illustrated in Fig.~\ref{fig:fig1} with the example of the open Rabi model (see Refs.~\cite{PhysRevA.97.013825} and below), a paradigmatic model for the study of dissipative phase transitions with finite components. Figure~\ref{fig:fig1}(a) presents a schematic of our sensing protocol. We assume that the open system is tuned to a dissipative CP. The quantity we wish to sense is encoded as a parameter, $\theta$, in the system Hamiltonian. A single sensing interrogation consists of initializing the system at $t_0=0$ in the same (arbitrary) state and subsequent detection and evolution spanning $[0,t)$. We emphasize that such evolution is subjected to measurement backaction randomly driving the system away from the steady state, in stark contrast to the steady-state-based scenario~\cite{PhysRevA.96.013817,PhysRevLett.124.120504}. As such, our scheme has the natural advantage that it does not require steady-state preparation that typically suffers critical slowing down. Detection of the radiation quanta continuously in time provides us with measurement signals $D(t,0)$ dependent on the system dynamics, that is, on $\theta$. This allows us to construct an estimator $\theta_{\rm est}$.
The associated precision can be quantified by the Fisher information (FI) of the detected signal
\begin{equation} \label{eq:FI0}
F_\theta(t) = \sum_{D(t,0)}P[D(t,0)]\left\{\partial_\theta {\rm{ln}} P[D(t,0)] \right\}^2,
\end{equation}
where $P[D(t,0)]$ is the probability of the continuously detected signal $D(t,0)$. According to the Cram\'er-Rao inequality~\cite{Kay97}, the FI sets a lower bound to the variance of any (unbiased) estimator of $\theta$, i.e., ${\rm Var}(\theta_{\rm est})\geq1/F_\theta(t)$.

The universal scaling of the precision bounds is exemplified in Fig.~\ref{fig:fig1}(b) by the sensing of the cavity mode frequency, $\theta=\omega$. We find that the time dependence of both the global QFI and the FI can be divided into two regimes, the transient regime and the long-time regime, separated by a characteristic time scale $t_*\simeq\eta/\kappa $, with $\eta$ the effective size of the system and $\kappa$ the cavity damping rate.
In the transient regime, $t\lesssim t_*$, the global QFI obeys a super-Heisenberg scaling with respect to time $I_\omega(t)\sim t^3$ whereas the FI obeys the Heisenberg scaling $F_\omega(t)\sim t^2$. In the long-time regime, $t \gg t_*$, both quantities grow linearly in time, and depend algebraically on the system size $\eta$, $I_\omega(t) \sim  \eta^2 t$ and $F_\omega(t) \sim \eta t$. The scaling exponents of the global QFI can be {expressed analytically} in terms of the critical exponents of the underlying dissipative CP.

The rest of the paper is organized as follows. In Sec.~\ref{sec: global_QFI}, we develop a scaling theory for the universal behavior of the global QFI~\eqref{eq: gQFI_1} at a generic dissipative critical point. In Sec.~\ref{sec:continous_measure}, we analyze a continuous-measurement-based sensing protocol for accessing the criticality-enhanced precision, illustrated with the example of the open Rabi model. We conclude in Sec.~\ref{sec: conclusion} with a summary of our results and an outlook.

\section{UNIVERSAL SCALING OF THE GLOBAL QUANTUM FISHER INFORMATION}
\label{sec: global_QFI}
Let us start our discussion by analyzing the precision limit of a generic quantum-optical open sensor at dissipative criticality. Consider the sensor as an open many-particle system with spatial extension $L$ defined on a $d$-dimensional lattice, consisting of $N=L^d$ interacting particles as individual sensor components. We make the typical assumptions underlying quantum-optical master equations (see e.g., Chapter 3 of Ref.~\cite{breuer2002theory}), i.e., the open system is weakly coupled to a Markovian environment at zero temperature. Consequently, the reduced density matrix of the unobserved system evolves according to a Lindblad master equation (LME)~\cite{carmichael1993open,breuer2002theory,gardiner2004quantum,wiseman_milburn_2009,Rivas_2012} (we set $\hbar=1$ hereafter)
\begin{equation}
\label{eq:LME}
\dot{\rho}={\cal L} \rho \equiv -i [\hat{H}(\theta),\rho] 
+\sum _{\ell} \left( \hat{J}_{\ell} \rho \hat{J}_{\ell}^\dag 
-\frac{1}{2}\{\hat{J}_\ell^\dag\hat{J}_\ell, \rho\} \right).
\end{equation}
We consider the sensing of a single quantity that, without the loss of generality, is assumed to be encoded as a parameter $\theta$ of the many-body Hamiltonian $\hat{H}(\theta)$. Here $\{\hat{J}_{\ell}\}$ represents a set of (single- or many-body) jump operators resulting from system and environment coupling, with $\ell$ the set index. To be specific, we focus on the experimentally common situation where $\theta$ is encoded in single-body terms,
\begin{equation}
\label{eq: assume_h}
\hat{H}(\theta) = \sum_{i=1}^N \hat{h}_i(\theta)+\sum_{i<j}^N\hat{h}_{ij}+\dots,
\end{equation}
i.e., we assume that all $k$-body ($k\geq2$) terms describing inter-particle interactions are $\theta$ independent.

The sensor precision is upper bounded by the global QFI~\eqref{eq: gQFI_1}, which may be achieved by the most general measurement performed on the joint system and environment state $|\Psi(t)\rangle$. For quantum-optical open sensors where the quantum regression formula~\cite{1998aibp,gardiner2004quantum} typically holds, the global QFI can be expressed in terms of the system autocorrelators~\cite{PhysRevLett.112.170401}, 
\begin{equation}
\label{Moriginal}
I_\theta(t)= 2 \int_0^t d\tau \int_0^t d\tau' \langle \{ \delta \hat{O}(\tau'),\delta \hat{O}(\tau) \} \rangle.
\end{equation}
Here, $\hat{O}:=\partial_\theta \hat{H}(\theta)$ is the Hermitian operator which encodes $\theta$, the angle bracket $\langle \cdots\rangle:={\rm tr}[\cdots\rho(0)]$ denotes an expectation with respect to the initial (pure) system density matrix $\rho(0)=|\psi(0)\rangle\langle\psi(0)|$, and $\delta \hat{O}(t):=\hat{O}(t) -\langle \hat{O}(t) \rangle$. Under our assumption Eq.~\eqref{eq: assume_h}, $\hat{O}=\sum_{i=1}^{N}\hat{o}_i$ is an extensive single-body operator, with $\hat{o}_i:=\partial_\theta \hat{h}_i(\theta)$ the local operator of the $i$-th particle. The global QFI \eqref{Moriginal} can therefore be expressed as
\begin{align}
I_\theta(t,L)&= 8L^d\int_0^t d\tau\int_0^{t-\tau} ds \,S(\tau,s,L),\label{eq:QFI_correlators} \\
S(\tau,s,L)&=\frac{1}{L^d}\sum_{i,j=1}^N \mathfrak{R}[\langle \delta \hat{o}_i(\tau+s)\delta \hat{o}_j(\tau) \rangle],\label{eq:dynamic_structure_factor} 
\end{align}
in which $\mathfrak{R}$ denotes the real part, and we have indicated explicitly the dependence of the global QFI on the system size $L$. Equations \eqref{eq:QFI_correlators} and \eqref{eq:dynamic_structure_factor} provide us with a useful connection between the global QFI and the autocorrelators of the open quantum sensor. In particular, as we show below, they lead to explicit quantitative predictions at continuous dissipative CPs, where the universal scaling laws of the autocorrelators translate directly into the scaling of the global QFI.
\subsection{Criticality and universal scaling}
\label{sec:Criticality and universal scaling}
Let us {assume} that LME \eqref{eq:LME} supports a continuous dissipative CP in its steady state $\rho_{\rm st}$, i.e., the solution of the stationary LME ${\cal{L}} \rho_{\rm st}=0$. We assume that $h$ is a parameter of LME \eqref{eq:LME} that drives the system across the phase transition ($h$ can be independent of $\theta$), and the CP is located at $h_{\rm CP}=0$. Close to the CP, the static and dynamic properties associated with $\rho_{\rm st}$ obey universal scaling laws governed by a small number of critical exponents. In the following we show that the global QFI~\eqref{eq:QFI_correlators} manifests such universal scaling behavior. 

To begin with, we introduce
\begin{equation*}
S_{\rm st}(s,L)=\frac{1}{L^d} \sum_{i,j=1}^N\mathfrak{R}[\langle \delta \hat{o}_i(s)\delta \hat{o}_j(0) \rangle_{\rm st}],
\end{equation*}
with $\langle\cdots\rangle_{\rm st}:={\rm tr}(\rho_{\rm st}\cdots)$ an expectation value with respect to the steady state. Note that $S_{\rm st}(0,L)$ is the static structure factor of the steady state at zero momentum. Assuming translational invariance, and taking the thermodynamic limit $L\to \infty$, $S_{\rm st}(0,L)$ obeys the well-known scaling behavior (see, e.g., Ref.~\cite{cardy}) $S_{\rm st}(0,L)\sim \xi^{d-2\Delta_{\hat{o}}}$, provided $d-2\Delta_{\hat{o}}>0$, where $\Delta_{\hat{o}}$ is the scaling dimension of the operator $\hat{o}_i$, defined via $\langle \delta\hat{o}_i(0)\delta\hat{o}_{i+r}(0)\rangle_{\rm st}\sim  r^{-2\Delta_{\hat{o}}}$ at the CP $h_{\rm CP}=0$; $\xi$ is the correlation length of the system, which diverges according to $\xi\sim h^{-\nu}$ close to the CP, with $\nu$ the correlation-length critical exponent. In the opposite case {of} $d-2\Delta_{\hat{o}}\leq 0$, $S_{\rm st}(0,L)\sim {\rm constant}$ dependent on the short-distance (ultraviolet) cutoff, i.e., it is not universal. For correlations of nonequal time $s\neq 0$ and of finite system size $L$, the above scaling behavior can be generalized to a dynamic scaling form~\cite{PhysRevResearch.2.023211,PhysRevE.97.052148}
\begin{equation}
\label{eq:S_st_scaling}
S_{\rm st}(s,L)=\Lambda^{d-2\Delta_{\hat{o}}}\phi_{\rm st}(\Lambda^{1/\nu}h, \Lambda^z s^{-1},\Lambda L^{-1}),
\end{equation}
in which $\Lambda$ is a cutoff length scale determined by the relative strength of the two relevant perturbations to the CP, $h$ and $L^{-1}$, $\phi_{\rm st}$ is a universal scaling function and $z$ is the dynamic critical exponent. Equation~\eqref{eq:S_st_scaling} completely captures the dominant divergent behavior of $S_{\rm st}(s,L)$ close to criticality, i.e., for small enough $h$ and $L^{-1}$. If $h$ is sufficiently small, $h\ll L^{-1/\nu}$, the inverse system size $L^{-1}$ is the dominant perturbation to the CP, i.e., the finite-size effect most severely drives the system away from criticality. Correspondingly, $\Lambda\simeq L$ and therefore $S_{\rm st}(s,L)\sim L^{d-2\Delta_{\hat{o}}}$. In the opposite limit $h\gg L^{-1/\nu}$, $h$ is the dominant perturbation to the CP. Correspondingly, $\Lambda\simeq \xi\sim h^{-\nu}$ and thus $S_{\rm st}(s,L)\sim h^{-\nu d+2\nu\Delta_{\hat{o}}}$.

The scaling behavior of $S(\tau,s,L)$ [cf. Eq.~\eqref{eq:dynamic_structure_factor}] can be deduced via direct extensions of Eq.~\eqref{eq:S_st_scaling}. To keep our presentation concise, we only outline the essential physics here, and defer a comprehensive theoretical analysis to Appendix~\ref{appA}. We note that in the limit $\tau\to \infty$, the auto-correlator $S({\tau,s,L})\to S_{\rm st}({s,L})$, as can be seen by applying the quantum regression formula $\langle \delta\hat{o}_i(\tau+s) \delta\hat{o}_j(\tau)\rangle={\rm tr}[\delta\hat{o}_i e^{{\cal L} s}\delta\hat{o}_j e^{\cal L\tau}\rho(0)]$. Consequently, the parameter $\tau^{-1}$ can be regarded as a new perturbation that controls the scaling behavior of $S(\tau,s,L)$ close to criticality, i.e., for $\tau\gg 1$. Importantly, $\tau^{-1}$ can be shown to be a relevant perturbation (in the renormalization group sense) with a scaling dimension identical to the dynamic exponent $z$. This allows us to extend Eq.~\eqref{eq:S_st_scaling} by the inclusion of $\tau^{-1}$ as a new scaling variable besides $h$ and $L^{-1}$, therefore arriving at a dynamic scaling ansatz 
\begin{equation} 
\label{scaling_correlators}
S(\tau,s,L)= \Lambda^{d-2 \Delta_{\hat{o}}} \phi(\Lambda^{1/\nu}h, \Lambda^z \tau^{-1}, \Lambda^z s^{-1},\Lambda L^{-1}),
\end{equation}
where the cutoff length scale $\Lambda$ is determined by the relative strengths of $h$, $\tau^{-1}$ and $L^{-1}$. We emphasize that Eq.~\eqref{scaling_correlators} is valid in the critical region, where the relevant perturbations $h$, $\tau^{-1}$ and $L^{-1}$ are sufficiently small such that the system behavior is universal. In particular, the evolution time $\tau$ in Eq.~\eqref{scaling_correlators} should be longer than some short cutoff time scale $t_{\rm cutoff}$, whose exact value depends on the model details (i.e., it is not universal) and is not interesting to us. To focus on the universal scaling of precision bounds, we implicitly assume that all time scales we consider are longer than $t_{\rm cutoff}$ in the rest of the paper.

We provide a comprehensive formulation of the physics outlined above in Appendix~\ref{appA} via analogy with quantum criticality based on the renormalization group framework. Remarkably, we show theoretically that the scaling ansatz \eqref{scaling_correlators} is applicable to generic continuous dissipative CPs, except the rare case of multicritical points. To consolidate our theoretical analysis, we verify Eq.~\eqref{scaling_correlators} for the model system considered in our work (the open Rabi model, to be introduced in Sec.~\ref{sec:Quantum_Rabi}) via numerical finite-size scaling, which can be extended to verify Eq.~\eqref{scaling_correlators} for other models as well.

Equation \eqref{scaling_correlators} provides us with a simple yet powerful scaling form for autocorrelators close to dissipative criticality, which leads directly to the scaling laws of the global QFI via Eq.~\eqref{eq:QFI_correlators}, as shown in the next section.

\subsection{Scaling laws of the global quantum Fisher information}
\label{sec:scaling_law_global_QFI}
We now study the scaling laws of the global QFI \eqref{eq:QFI_correlators}, based on the scaling ansatz~\eqref{scaling_correlators}. First, we note that Eq.~\eqref{scaling_correlators} allows us to identify different regimes where the cutoff length scale $\Lambda$ is set by different perturbations to the CP among $h$, $L^{-1}$ and $\tau^{-1}$, and correspondingly $S(\tau,s,L)$ obeys different scaling forms. For criticality-enhanced sensing, we consider $h=0$, i.e., the system parameter is tuned at the CP.  
Consequently, $S(\tau,s,L)$ picks up the following scaling forms
\begin{numcases} {\hspace{-5mm} S(\tau,s,L)=}
 \tau^{(d-2 \Delta_{\hat{o}})/z}  \phi_\tau(\tau s^{-1},\tau L^{-z}), & \hspace{-5 mm} $\tau\lesssim L^z$,    \label{eq:correlators_short_time}\\
L^{d-2 \Delta_{\hat{o}}} \phi_L(L^{z}\tau^{-1},L^{z}s^{-1}),  & \hspace{-5 mm} $\tau \gg L^z$,\label{eq:correlators_long_time}
\end{numcases}
in which $\phi_\tau$ and $\phi_L$ are universal scaling functions inherited from $\phi$ in Eq.~\eqref{scaling_correlators}.

Substituting these scaling forms into Eq.~\eqref{eq:QFI_correlators}, we see immediately that the global QFI manifests two distinct scaling behaviors dependent on the total interrogation time $t$ compared to the system size $L$: (i) $t\lesssim L^z$, which is referred to as the transient regime hereafter, where the integrand $S(\tau,s,L)$ obeys the scaling form~\eqref{eq:correlators_short_time}; (ii) $t\geq L^z$, which is referred to as the long-time regime hereafter, where $S(\tau,s,L)$ picks the form~\eqref{eq:correlators_long_time} in most of the integration interval in Eq.~\eqref{eq:QFI_correlators}. Completing the integration in Eq.~\eqref{eq:QFI_correlators} in the two regimes we find that
\begin{numcases}{I_\theta(t,L) \sim} 
t^{(d-2 \Delta_{\hat{o}})/z+2}L^d, &$t\lesssim L^z$,\label{eq:QFI_short_time_general} \\
t L^{2d-2 \Delta_{\hat{o}}+z}, & $t \gg L^z$.\label{eq:QFI_long_time_general}
\end{numcases}
In arriving at Eq.~\eqref{eq:QFI_long_time_general}, we have made the approximation $\int_0^{t-\tau} ds \simeq\int_0^{\infty} ds$, justified by the fact that $\phi_L(L^{z}\tau^{-1},L^{z}s^{-1})\to 0$ sufficiently fast at large $s/L^z$ as a result of the finite correlation time in a finite-size system.

Equations \eqref{eq:QFI_short_time_general} and \eqref{eq:QFI_long_time_general} serve as the central formulas of this section, which provide us with the scaling laws of the global QFI \eqref{eq: gQFI_1} of a generic quantum-optical open quantum sensor at dissipative criticality. Such scaling depends on the spatial dimension $d$, the dynamic critical exponent $z$ and the scaling dimension of the local operator $\hat{o}_i$ that encodes the unknown parameter.

\subsection{Super-Heisenberg scaling and consistency with the Heisenberg limit}
\begin{figure}[b!]
\centering{}
\includegraphics[width=0.48\textwidth]{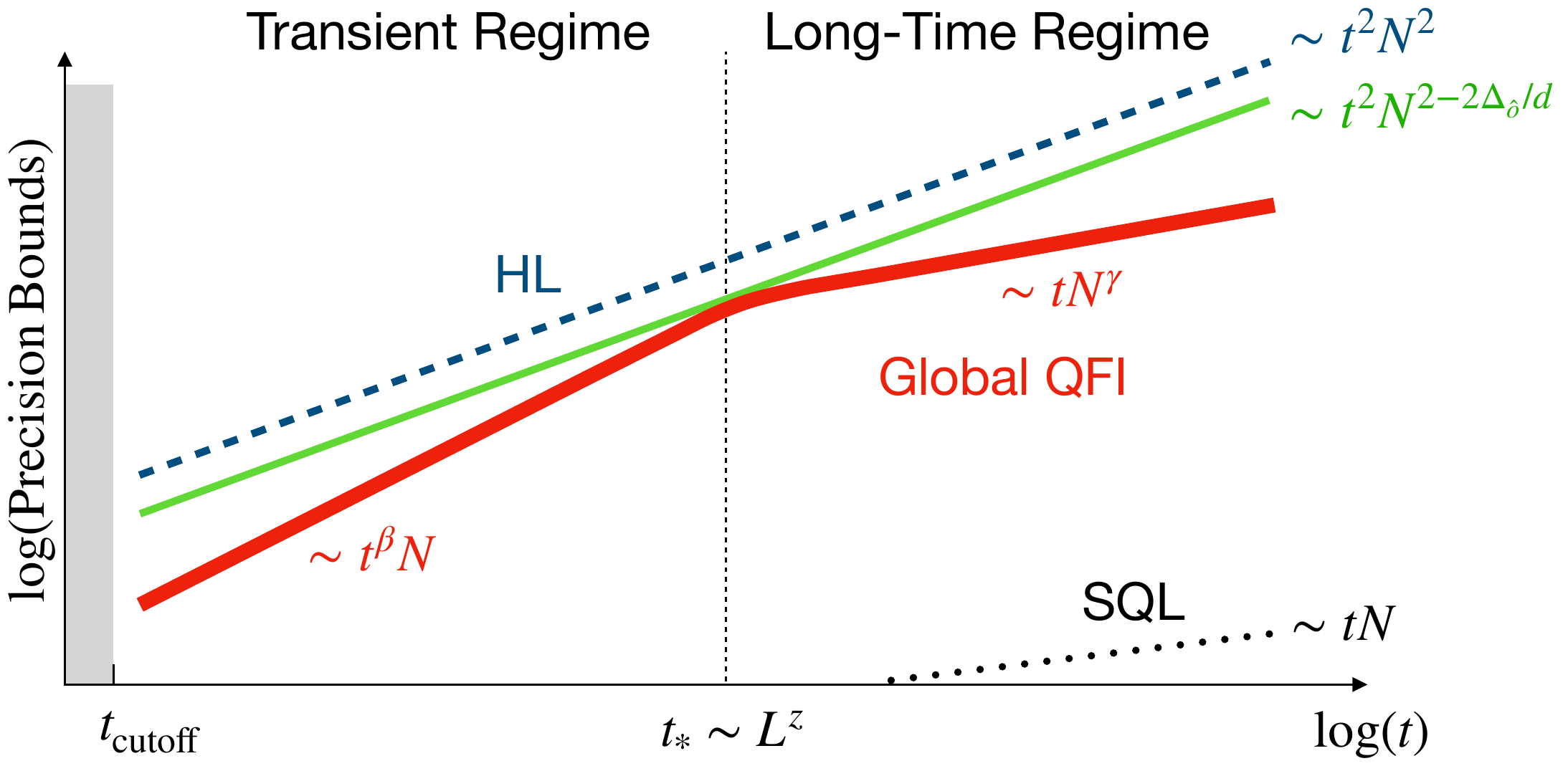} 
\caption{Illustration of the universal scaling laws of the global QFI at a generic dissipative CP, Eqs.~\eqref{eq:QFI_short_time_general} and \eqref{eq:QFI_long_time_general}, with respect to the interrogation time $t$ and the particle number $N\equiv L^d$, assuming $d\geq 1$. The global QFI (red solid line) and its tight bound Eq.~\eqref{eq: practical_bound} (green solid line), are contrasted with the HL (blue dashed line) and the SQL (black doted light). The characteristic time scale $t_*\sim L^z$ separates the transient regime and the long-time regime, where the global QFI obeys different scaling laws. The scaling exponents are determined by the spatial dimension $d$, the dynamic critical exponent $z$ and the scaling dimension $\Delta_{\hat{o}}$ of the local operator $\hat{o}_i$ that encodes the unknown parameter, via $\beta=2+(d-2\Delta_{\hat{o}})/z$ and $\gamma=2+(z-2\Delta_{\hat{o}})/d$. 
The shaded region represents an initial time window characterized by a short cutoff time scale $t_{\rm cutoff}$, in which the behavior of the global QFI is not universal.
}
\label{fig:fig1.5} 
\end{figure}

Let us discuss a few peculiar features of the scaling laws in Eqs.~\eqref{eq:QFI_short_time_general} and \eqref{eq:QFI_long_time_general}; cf. Fig.~\ref{fig:fig1.5}. First, we note that {they} surpass the SQL, and are improved further by coupling $\theta$ to operators with a small scaling dimension $\Delta_{\hat{o}}$. Such a condition is typically met by relevant operators (in the renormalization group sense) of low-dimensional CPs~\cite{PhysRevX.8.021022}.
Second, we compare the scaling laws in Eqs.~\eqref{eq:QFI_short_time_general} and \eqref{eq:QFI_long_time_general} to the HL widely discussed in the context of interferometric sensing~\cite{PhysRevD.23.1693,Giovannetti1330}. We focus on the case that $\hat{o}_i$ is a single-body operator, and we temporarily assume $d\geq 1$. Therefore, the HL adopts the familiar expression $\mathrm{H L}\sim t^2N^2\equiv t^2 L^{2d}$. We comment on the special case $d=0$ at the end of this section. 

One interesting aspect is the apparent super-Heisenberg scaling with respect to the interrogation time $t$ or the particle number $N$ in different regimes. In the transient regime $t\lesssim L^z$, the global QFI [cf. Eq.~\eqref{eq:QFI_short_time_general}] manifests super-Heisenberg scaling with respect to $t$ [note that we assume that $d-2\Delta_{\hat{o}}>0$ for the validity of Eqs.~\eqref{eq:QFI_short_time_general} and \eqref{eq:QFI_long_time_general}] and linear scaling with respect to $N$. In the long-time regime $t\gg L^z$, the global QFI [cf. Eq.~\eqref{eq:QFI_long_time_general}] manifests linear scaling in $t$ {and} super-Heisenberg scaling in $N$, provided by the condition $z-2\Delta_{\hat{o}}>0$, which can be fulfilled by encoding $\theta$ in operators with small scaling dimensions.

Despite such apparent super-Heisenberg scaling, the global QFI is indeed upper-bounded by the HL. To see this, let us introduce the precision scaling 
\begin{equation}
\label{eq: practical_bound}
I_{\theta}^{*}(t,L)\sim t^2L^{2d-2\Delta_{\hat{o}}}.
\end{equation}
We note that for continuous CPs defined in spatial dimensions $d\geq 1$, the operator scaling dimensions are positive, $\Delta_{\hat{o}}\geq0$. Therefore, scaling \eqref{eq: practical_bound} is sub-Heisenberg. Dividing both sides of Eqs.~\eqref{eq:QFI_short_time_general} and \eqref{eq:QFI_long_time_general} by $I_{\theta}^{*}(t,L)$ yields
\begin{equation*}\label{eq:QFI_ratio}
{I_\theta(t,L)}/{I_{\theta}^{*}(t,L)} \sim
\begin{cases}
(t L^{-z})^{(d-2\Delta_{\hat{o}})/z}, &t\lesssim L^z \\
{L^z}{t^{-1}}, & t \gg L^z.
\end{cases}
\end{equation*}
Therefore, $I_{\theta}(t,L)\lesssim I_{\theta}^*(t,L)$ in both the transient and the long-time regimes. Only at the sweet spot $t_*\simeq L^z$, $I_{\theta}(t_*,L)$ saturates $I_\theta^*({t,L})$. This confirms that the global QFI is Heisenberg limited in $d\geq1$. 

In $d=0$, the critical open system does not possess a rigorous spatial extension $L$. Consequently, the HL cannot be expressed as $t^2L^{2d}$, and should be analyzed in terms of the actual resources involved in the sensing protocol. This is illustrated in the next section via the open Rabi model, for which we show that the associated global QFI can actually saturate the HL [cf., Fig.~\ref{fig:fig1}(b) and Sec.~\ref{sec:QFI_Rabi}]. 

\section{CRITICALITY-ENHANCED PRECISION VIA CONTINUOUS MEASUREMENT}
\label{sec:continous_measure}
The universal scaling of the global QFI opens an attractive avenue towards enhanced precision scaling by harnessing dissipative criticality. Saturating the global QFI, however, requires a most general measurement of the joint system and environment, which may be practically challenging. This leads to the question of whether practical measurement schemes provide access to the criticality-enhanced scaling (which, in general, is lower than the scaling of the global QFI).
In this section we provide a positive answer to this question, by {analyzing} a readily implementable sensing protocol based on continuous measurement of the radiation quanta emitted by the critical open sensor. While our protocol is generally applicable to critical open systems in diverse setups that permit optical readout, to be specific in the following we illustrate it via the open Rabi model~\cite{PhysRevA.97.013825}, a light-matter interacting model featuring a continuous dissipative CP in zero {spatial} dimension. Besides its conceptual simplicity, its finite-component nature facilitates numerical simulation, making it possible to extract the scaling of the (classical) Fisher information as the precision bound of our sensing scheme. As such, it allows for direct comparison between the precision scaling of our protocol and that of a recent study~\cite{PhysRevLett.124.120504} based on direct, instantaneous measurement of the critical steady state of the open Rabi model, therefore demonstrating the advantages of our scheme.

\subsection{Model system: the open Rabi model}
\label{sec:Quantum_Rabi}
Let us consider a cavity mode coupled to a qubit [cf. Fig.~\ref{fig:fig1}(a)] as described by the quantum Rabi Hamiltonian
\begin{equation} \label{eq: Hamiltonian}
\hat{H}=\omega \hadag \ha +\frac{\Omega}{2} \hat{\sigma}_z-\lambda  (\ha+\hadag)\hat{\sigma}_x.
\end{equation} 
Here, $\ha$ $(\hadag)$ denotes the annihilation (creation) operator of the cavity mode, $\hat{\sigma}_{z,x}$ are the Pauli matrices of the qubit, $\omega$ is the cavity mode frequency, $\Omega$ is the qubit transition frequency and $\lambda$ is the coupling strength. The cavity leaks photons to the external electromagnetic environment at a rate $\kappa$. Without monitoring the environment, the dynamics of the open cavity-qubit system can be described by a standard LME~\cite{PhysRevA.97.013825,PhysRevLett.124.120504}
\begin{equation}
\label{eq:LME_Rabi}
\dot{\rho}= -i [\hat{H},\rho] 
+\kappa\left( \hat{c} \rho \hat{c}^\dag 
-\frac{1}{2}\{\hat{c}^\dag\hat{c}, \rho\} \right).
\end{equation}
Equation \eqref{eq:LME_Rabi} conserves a ${\mathbb Z}_2$ parity symmetry $\hat{c}\to-\hat{c},\hat{\sigma}_x\to-\hat{\sigma}_x$. Remarkably, in the soft-mode limit ${\omega}/\Omega\to 0$, the steady state of Eq.~\eqref{eq:LME_Rabi} can spontaneously break such a symmetry, resulting in a continuous dissipative phase transition~\cite{PhysRevA.97.013825}. 

Being zero dimensional, the model \eqref{eq:LME_Rabi} does not possess a rigorously defined system size. Nevertheless, following Refs.~\cite{PhysRevA.97.013825,PhysRevLett.115.180404} we can introduce the frequency ratio $\eta=\Omega/{\omega}$ as the `effective size' of the model, with $\eta\to\infty$ corresponding to the thermodynamic limit~\cite{PhysRevA.97.013825}. We further introduce a dimensionless coupling constant $g=2 \lambda/\sqrt{\Omega \omega}$. In the limit $\eta \rightarrow \infty$, the system undergoes a continuous phase transition at a CP $g_{\rm CP}=\sqrt{1+(\kappa/2\omega)^2}$~\cite{PhysRevA.97.013825}, cf. Fig.~\ref{fig:fig1}(a). At $g<g_{ \rm CP}$, the system is in the normal phase, characterized by the order parameter $\langle\hat{c}\rangle_{\rm st}=0$, whereas at $g>g_{ \rm CP}$, the system enters the superradiant phase, characterized by $\langle\hat{c}\rangle_{\rm st}\neq 0$. Finite $\eta$ plays the role of a finite-size cutoff, which renders the phase transition to a smooth crossover.

The critical properties of the CP have been extracted numerically via finite-size(i.e., finite-$\eta$) scaling of various thermodynamic quantities~\cite{PhysRevA.97.013825}, which provides $z=1$ as the dynamic and $\nu=2$ as the correlation length critical exponent. As such, the open Rabi model lies in the same universality class as the open Dicke model~\cite{PhysRevA.75.013804, PhysRevLett.104.130401}. At the CP $g=g_{\rm CP}$, the mean occupation of the cavity mode diverges according to $\langle\hat{n}\rangle_{\rm st}\sim \eta^{1/2}$, in which $\hat{n}:=\hat{c}^\dag\hat{c}$. Therefore, the scaling dimension of $\hat{n}$ is $\Delta_{\hat{n}}=-1/2$, which satisfies the condition $d-2\Delta_{\hat{n}}>0$ (cf. Sec.~\ref{sec:scaling_law_global_QFI}). As such, the sensing of the cavity mode frequency $\omega$ serves as an ideal demonstration of criticality-enhanced sensing, which we analyze in detail in sections below.

Finally, we emphasize that the features of the open Rabi model presented above fundamentally rely on the validity of LME \eqref{eq:LME_Rabi} in the thermodynamic limit $\eta\to\infty$, as assumed without proof by previous studies~\cite{PhysRevA.97.013825,PhysRevLett.124.120504}. In the present work we manage to take the significant leap of a rigorous derivation of Eq.~\eqref{eq:LME_Rabi} in such a limit, based on the microscopic system and environment Hamiltonian in an actual experimental implementation. We outline such an implementation in Sec.~\ref{sec:Experimental_implementation}, and provide the detailed derivation in Appendix \ref{appD}, which consolidate the model system for demonstrating our sensing scheme and provide a firm ground for Refs.~\cite{PhysRevA.97.013825,PhysRevLett.124.120504}.

\subsection{Scaling laws of the global quantum Fisher information}
\label{sec:QFI_Rabi}
\begin{figure}[t!]
\centering{}
\includegraphics[width=0.48\textwidth]{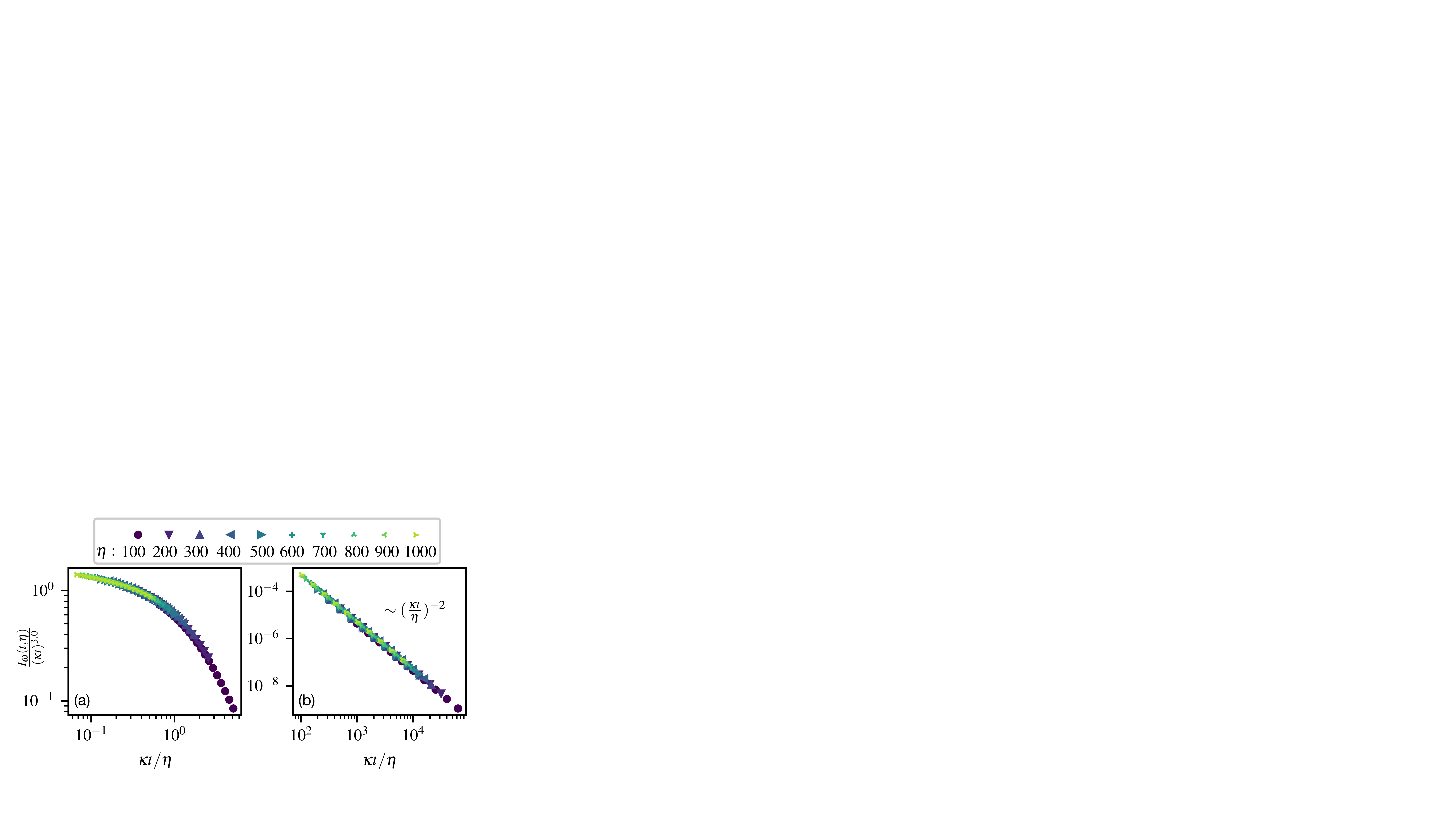} 
\caption{Finite-size scaling of the global QFI of the critical open Rabi model plus the environment, in both (a) the transient regime $\kappa t\lesssim \eta$ and (b) the long-time regime $\kappa t\gg \eta$. The open system is at a dissipative critical point $g=g_{\rm CP}$, and $\eta$ is its effective size (see text). Parameters: $\kappa=0.1$, $\omega=1$.}
\label{fig:fig2} 
\end{figure}
In Fig.~\ref{fig:fig2} we show the numerical finite-size scaling of the global QFI for the sensing of $\omega$, for different system sizes $\eta$ in both the transient regime $\kappa t\lesssim\eta^z$ [Fig.~\ref{fig:fig2}(a)] and in the long-time regime $\kappa t\gg\eta^z$ [Fig.~\ref{fig:fig2}(b)], assuming that the model is tuned to the CP $g=g_{\rm CP}$. The perfect data collapse indicates the following scaling behavior of the global QFI
\begin{numcases}{I_\omega(t,\eta)=} 
(\kappa t)^3 f_I\left(\kappa t/\eta\right), &$\kappa t\lesssim \eta$, \label{eq:QFI_short_time}\\
{\rm const.} \times\kappa t \eta^2,&$\kappa t \gg \eta$, \label{eq:QFI_long_time}
\end{numcases}
with $f_I(\kappa t/\eta)$ is a scaling function reflecting the finite-size correction. This validates the predictions of the general formulas Eqs. \eqref{eq:QFI_short_time_general} and \eqref{eq:QFI_long_time_general} when the relevant exponents $d=0, z=1$ and $\Delta_{\hat{n}}=-1/2$ are plugged in.

The scaling laws Eqs. \eqref{eq:QFI_short_time} and \eqref{eq:QFI_long_time} are Heisenberg limited. As a zero-dimensional model, the HL of the open Rabi model can be analyzed in terms of the actual resources involved, that is, the interrogation time $t$ and the mean occupation of the cavity mode $\langle\hat{n}\rangle_{\rm st}$. Hereby ${\rm HL}\sim t^2\langle\hat{n}\rangle_{\rm st}^2 \sim t^2\eta$. From Eqs.~\eqref{eq:QFI_short_time} and \eqref{eq:QFI_long_time}, we have $I_\omega(t,\eta)\sim (\kappa t/\eta)\times{\rm HL}$ in the transient regime $\kappa t\lesssim \eta$, whereas $I_\omega(t,\eta)\sim (\eta/\kappa t)\times{\rm HL}$ in the long-time regime $\kappa t\gg\eta$. Therefore, the global QFI manifests super-Heisenberg scaling with respect to the interrogation time $t$(the cavity photon number $\langle\hat{n}\rangle_{\rm st}$) in the transient(long-time) regime; nevertheless is always upper bounded by the HL. We further note that $I_\omega(t_*,\eta)\sim {\rm HL}$ at $t_*=\eta/\kappa$, i.e., the global QFI saturates the HL in the asymptotic limit $t,\eta\to\infty$ with $\kappa t\simeq \eta$ kept fixed. These features are illustrated in Fig.~\ref{fig:fig1}(b).

In contrast to these criticality-enhanced, Heisenberg-limited scaling laws at the CP, the global QFI obeys significantly lower scaling laws when the model is tuned far away from the CP. A detailed analysis of the latter is provided in Appendix \ref{appC}.

\subsection{Photon counting and the scaling laws of the Fisher Information}
Let us now analyze a continuous-measurement-based sensing protocol for accessing the criticality-enhanced precision scaling with open quantum sensors. As a concrete illustration, we consider the sensing of the cavity mode frequency $\omega$ of the open Rabi model via photon counting. Generalization to other types of continuous measurement, e.g., homodyning~\cite{PhysRevA.94.032103}, is straightforward.

We assume that the photons leaked from the cavity are directed to and counted by a photon detector [cf. Fig.~\ref{fig:fig1}(a)]. For simplicity, we assume unit detection efficiency (finite detection efficiency reduces the FI by an overall factor but does not change its scaling; see~\cite{Theodore}). The evolution of the joint cavity-qubit system is therefore subjected to the measurement backaction {conditioned} on a specific series of photon detection events. Specifically, in an infinitesimal time interval $d\tau$, the detection of a photon leads to the collapse of the (unnormalized) conditional state of the cavity-qubit system according to $|\tilde{\psi}_c\rangle\rightarrow \sqrt{\kappa d \tau}\ha |\tilde{\psi}_c\rangle$, while if no photon is detected, the system evolves according to the nonunitary dynamics $|\tilde{\psi}_c\rangle\rightarrow \hat{M}_0 |\tilde{\psi}_c\rangle$ with $\hat{M}_0=\hat{1}-d\tau(i \hat{H}+{\kappa} \hadag \ha/2)$~\cite{carmichael1993open,gardiner2004quantum,wiseman_milburn_2009,RevModPhys.70.101}. For an infinitesimal time interval $d\tau$, the probability of the detection of a photon is $p_1=\kappa \langle \hadag \ha \rangle_c d\tau$ with $\langle\cdots\rangle_c:=\langle\tilde{\psi}_c|\cdots|\tilde{\psi}_c\rangle/\langle\tilde{\psi}_c|\tilde{\psi}_c\rangle$, whereas the probability of the detection of no photon is $p_0=\langle\hat{M}_0^{\dagger} \hat{M}_0 \rangle_c=1-p_1$. Repeating such stochastic evolution for each time step $[\tau,\tau+d\tau)$, in which $\tau\in[0,t)$, defines a quantum trajectory consisting of the photon detection signals up to time $t$ and the associated conditional quantum state $|\tilde{\psi}_c(t)\rangle$.

Mathematically, this can be formulated rigorously in terms of a (It\^o) stochastic Schr\"odinger equation~\cite{carmichael1993open,gardiner2004quantum,wiseman_milburn_2009}
\begin{equation} 
\label{eq:sse_photon_counting}
d |\tilde{\psi}_c\rangle=-\left(i \hat{H} +\frac{\kappa}{2} \hadag \ha\right)dt|\tilde{\psi}_c\rangle + dN(t) (\sqrt{\kappa dt} \ha-\hat{1}) |\tilde{\psi}_c\rangle.
\end{equation} 
Here $dN(t)$ is a stochastic Poisson increment that takes two values: $dN(t)=0$ with probability $p_0$, and $dN(t)=1$ with probability $p_1$. The last term of Eq.~\eqref{eq:sse_photon_counting} accounts for the backaction of photon counting by updating the system state conditioned on the detection of a photon or not. Corresponding to Eq.~\eqref{eq:sse_photon_counting}, the photon-counting signal up to time $t$ for a specific trajectory is $D(t,0):=\{dN(n dt), ..., dN(dt), dN(0)\}$ where we identify $t\equiv ndt$. The probability of this trajectory is~\cite{carmichael1993open,gardiner2004quantum,wiseman_milburn_2009}
\begin{equation}
\label{eq:prob_traj}
P[D(t,0)]=p_{dN(n dt)} \cdots p_{dN(0)}= \langle \tilde{\psi}_c(t)| \tilde{\psi}_c (t) \rangle.
\end{equation}
An ensemble average over all conditional states leads to the definition of a density operator of the cavity-qubit system, $\rho(t)=\sum_{D(t,0)} |\tilde{\psi}_c(t)\rangle\langle\tilde{\psi}_c(t)|$, which evolves according to LME \eqref{eq:LME_Rabi}.

\begin{figure}[t!]
\centering{} \includegraphics[width=0.48\textwidth]{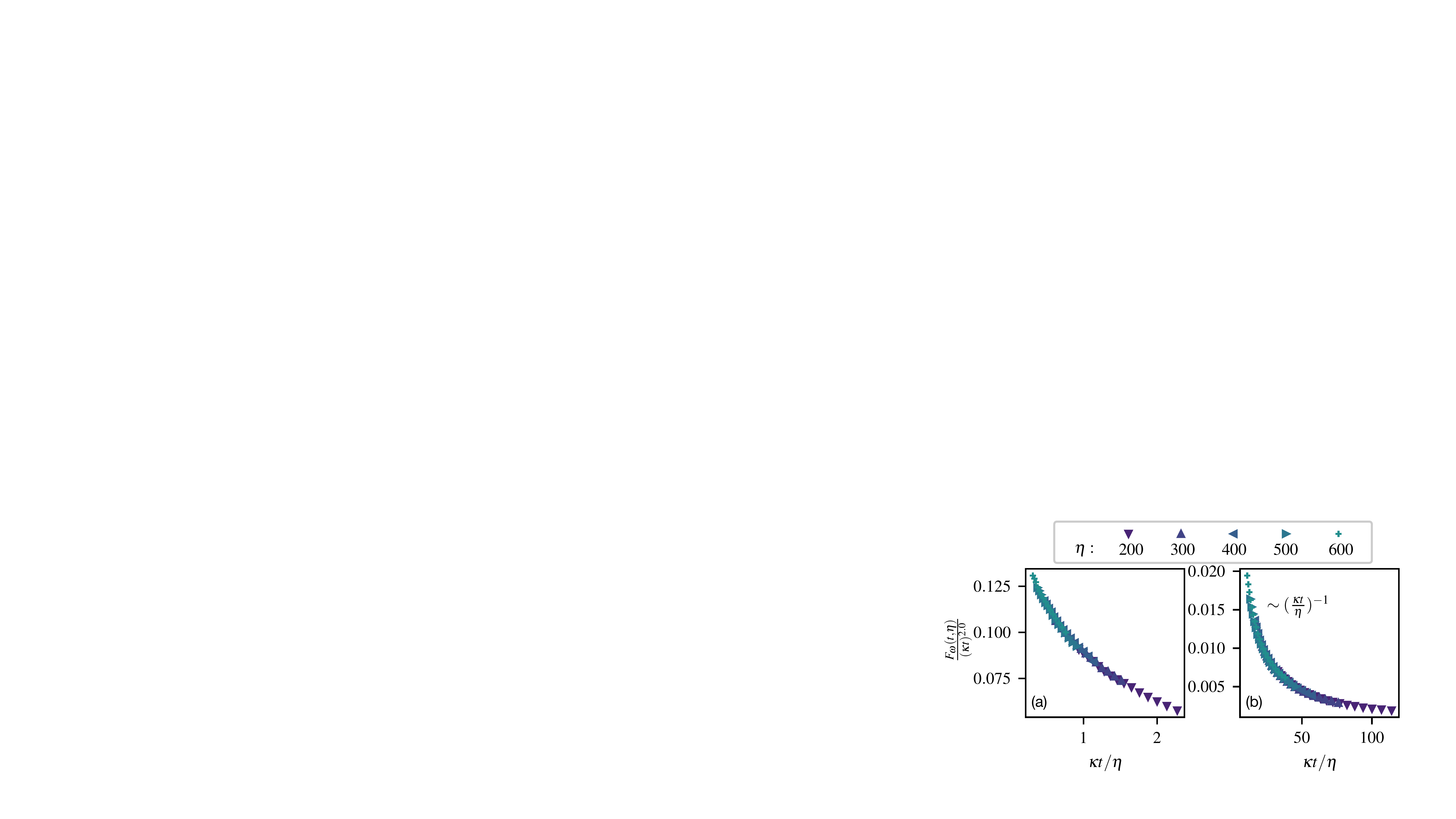} 
\caption{Finite-size scaling of the FI $F_\omega(t,\eta)$ [cf., Eq.~\eqref{eq:FI}], for the photon-counting signals of the critical open Rabi model ($g=g_{\rm CP}$), in both (a) the transient regime $\kappa t\lesssim \eta$ and (b) the long-time regime $\kappa t\gg \eta$. Parameters: $\kappa=0.1$, $\omega=1$. Each data point represents an average of $10^5$ independent trajectories in (a), whereas $10^4$ independent trajectories in (b). The sampling numbers are chosen such that the error associated with finite sampling is sufficiently small, cf. Fig.~\ref{fig:FI_trajectories}.}
\label{fig:fig3} 
\end{figure}
A single interrogation of our sensing protocol therefore corresponds to a quantum trajectory consisting of the continuously detected signal $D(t,0)$ and the associated conditional evolution of the open system. Processing $D(t,0)$ via standard means (e.g., via Bayesian inference~\cite{PhysRevA.87.032115,PhysRevA.87.032115,PhysRevA.89.052110,PhysRevA.94.032103}) provides an estimator $\omega_{\rm est}$ of the cavity mode frequency.
The associated precision is quantified by the FI of the detected signal
\begin{equation} \label{eq:FI}
F_\omega(t) = \sum_{D(t,0)}P[D(t,0)]\left\{\partial_\omega {\rm{ln}} P[D(t,0)] \right\}^2.
\end{equation}
FI \eqref{eq:FI} can be extracted numerically, via approximating the ensemble average $\sum_{D(t,0)}$ by a statistical average over sufficient (but finite) numbers of sampled trajectories. For each trajectory, we extract $P[D(t,0)]$ via Eq.~\eqref{eq:prob_traj} following the numerical propagation of Eq.~\eqref{eq:sse_photon_counting}, and $\partial_\omega P[D(t,0)]$ via calculating $P[D(t,0)]$ at a slightly different $\omega$ and subsequent numerical differentiation.

In Fig.~\ref{fig:fig3} we show the results of our numerical finite-size scaling of the FI for different system sizes $\eta$ in both the transient regime $\kappa t\lesssim\eta^z$ [Fig.~\ref{fig:fig3}(a)] and in the long-time regime $\kappa t\gg\eta^z$ [Fig.~\ref{fig:fig3}(b)], assuming that the model is tuned to the CP $g=g_{\rm CP}$. The perfect data collapse indicates, similar to the global QFI, that the FI obeys scaling behavior at criticality,
\begin{numcases}{F_\omega(t,\eta)}
 (\kappa t)^2 f_F(\kappa t/\eta), & $\kappa t\lesssim \eta$, \label{eq:FI_Rabi_transient}\\
{\rm const.} \times\kappa t \eta, & $\kappa t \gg \eta$.\label{eq:FI_Rabi_long_time}
\end{numcases}
where $f_F(\kappa t/\eta)$ is a scaling function reflecting the finite-size correction.
\begin{figure}[t!]
\centering{} \includegraphics[width=0.48\textwidth]{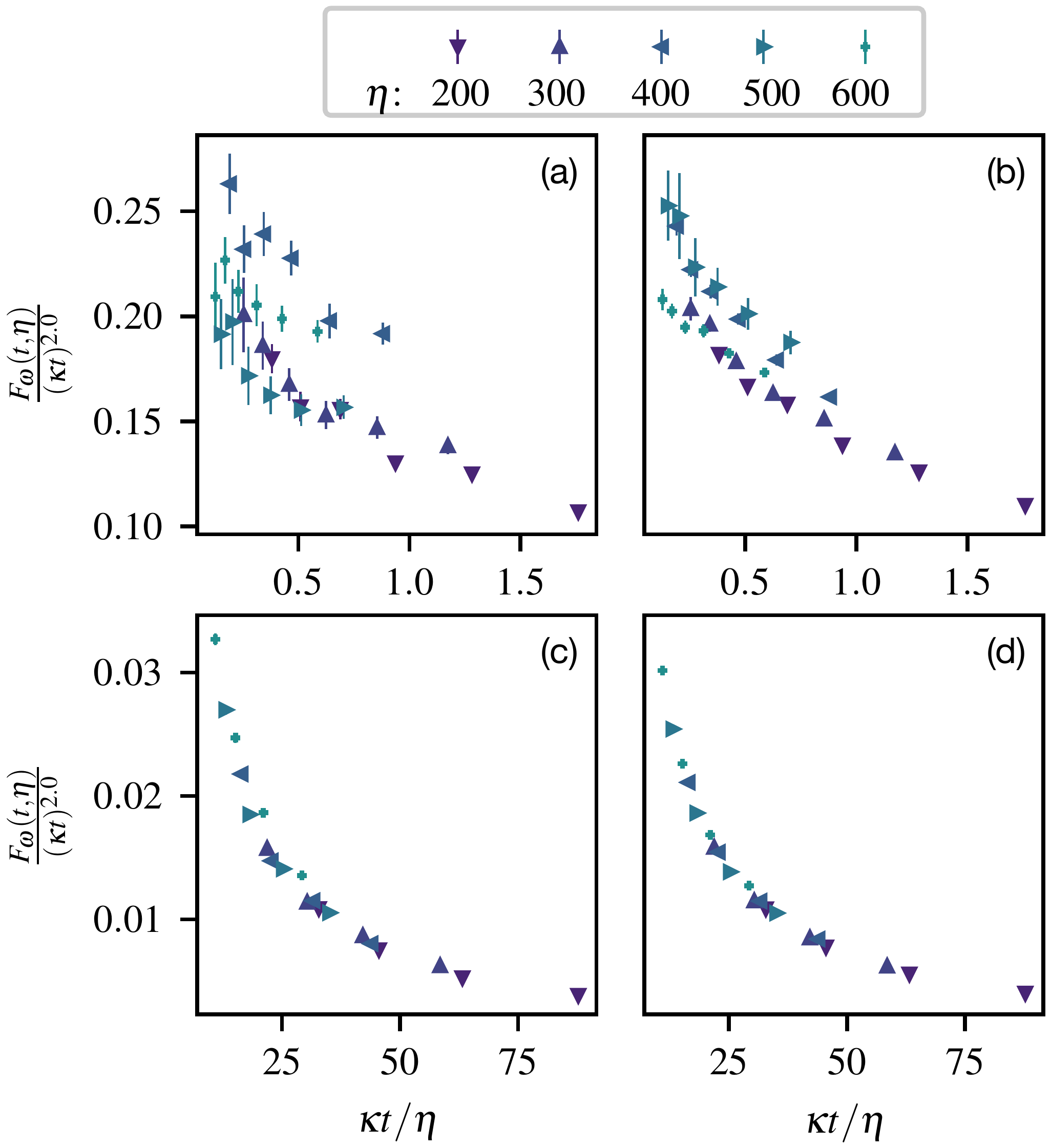} 
\caption{Convergence of FI~\eqref{eq:FI} with respect to the number of sampled trajectories $N_{\rm traj}$: (a),(c) $N_{\rm traj}=10^3$; (b),(d) $N_{\rm traj}=10^4$. Parameters are the same as Fig.~\ref{fig:fig3}. The error bars are estimated from the variance of ten independent samplings, and are not shown if they are smaller than the data point size. The FI converges more slowly in the transient regime [(a) and (b)] than in the long-time regime [(c) and (d)], as represented by larger error bars in the former. By sampling a sufficient number of trajectories we reduce the sampling error in the FI, leading to the scaling collapse shown in Fig.~\ref{fig:fig3}.}
\label{fig:FI_trajectories} 
\end{figure}

As a validation of the accuracy of our numerics, we show in Fig.~\ref{fig:FI_trajectories} the convergence of the approximated FI with respect to $N_{\rm traj}$, the number of sampled trajectories. As can be seen, the approximated FI gradually converges to the predicted scaling form (cf.~Fig.~\ref{fig:fig3}) as $N_{\rm traj}$ increases, while convincing data collapse typically requires $N_{\rm traj}\geq10^4$. As the simulation time of each trajectory scales polynomially in the system size at criticality, the extraction of such FI scaling for generic dissipative criticality in high spatial dimensions may represent a computational challenge. Here, thanks to its zero-dimensional, finite-component nature, our model system allows for extracting such scaling behavior directly by quantum trajectory simulation.

\subsection{Discussion}
\label{sec:physical_discussion_FI}
In contrast to the global QFI, we are not able to analytically relate the scaling laws of the FI given in Eqs. \eqref{eq:FI_Rabi_transient} and \eqref{eq:FI_Rabi_long_time} to the critical exponents of the underlying CP. Nevertheless, we manage to develop a physical understanding of these scaling laws by analyzing them with respect to the resources involved in our sensing protocol. First, the long-time behavior Eq.~\eqref{eq:FI_Rabi_long_time} can be expressed as $F_\omega(t,\eta)\sim \kappa^2 tt_*$ with $t_*\simeq \eta/\kappa$ as defined previously in Sec.~\ref{sec:QFI_Rabi}. This should be compared to the long-time behavior of the global QFI~\eqref{eq:QFI_long_time}, which can be expressed similarly as $I_\omega(t,\eta)\sim \kappa^2 tt_*\langle\hat{n}\rangle_{\rm st}^2$ (note that $\langle\hat{n}\rangle_{\rm st}\sim \eta^{1/2}$ at the CP, as introduced in Sec.~\ref{sec:Quantum_Rabi}). Such an expression illustrates that the global QFI \eqref{eq:QFI_long_time} behaves the same as that of a standard noiseless interferometric scheme involving $\langle\hat{n}\rangle_{\rm st}$ entangled photons and spanning a time window $t_*$ in each interrogation. The time scale $t_*$ therefore characterizes the correlation time of the emitted photons---two photon detection events that are separated by more than $t_*$ are essentially uncorrelated and are therefore analogous to two independent interrogations.  Such an interpretation extends naturally to the FI---in comparison to the global QFI, it lacks the contribution $\sim \langle\hat{n}\rangle_{\rm st}^2$ from the cavity photons, as it is obtained by measurement of the environment alone. 

In contrast to these long-time scalings, the transient scaling Eq. \eqref{eq:FI_Rabi_transient} is due to dynamic critical behavior that lacks an analog in conventional interferometric schemes. Nevertheless, it can be recovered from the long-time scaling Eq. \eqref{eq:FI_Rabi_long_time}  by the simple replacement $\eta\to (\kappa t)^{1/z}$, as they both originate from the dynamic criticality of the underlying CP, with $\eta^{-1}$ [$(\kappa t)^{-1/z}$] being the dominant perturbation in the long-time (transient) regime. Similarly for the global QFI, Eq.~\eqref{eq:QFI_short_time} can be related to~\eqref{eq:QFI_long_time} by the same replacement.

Finally, we compare the precision bounds of our continuous-measurement-based sensing protocol with that based on direct measurement of the critical steady state of the open Rabi model~\cite{PhysRevLett.124.120504}. The latter scheme discards the emitted radiation quanta and, therefore, the achievable sensing precision of it is upper bounded by the QFI of the reduced density matrix of the cavity-qubit system. Such a QFI was shown~\cite{PhysRevLett.124.120504} to obey the SQL, i.e., $\sim \kappa t \langle\hat{n}\rangle_{\rm st}\sim \kappa t\eta^{1/2}$, where the linear dependence on $t$ results from repeated preparation of the steady state in every interrogation. By continuously counting the emitted radiation quanta, the precision scaling is enhanced as quantified by the FI in the long-time regime $F_\omega(t,\eta)\sim \kappa t\eta$. Moreover, in our scheme the global QFI in the long-time regime obeys a further enhanced scaling $I_{\omega}(t,\eta)\sim \kappa t\eta^2$, which can in principle be achieved by a joint measurement of the open system and the environment.

\subsection{Experimental implementation} \label{sec:Experimental_implementation}
Various quantum optical platforms provide promising opportunities for the implementation of the model system demonstrating our sensing scheme. In particular, the Rabi Hamiltonian \eqref{eq: Hamiltonian} has been realized in trapped-ion~\cite{PhysRevX.8.021027,Cai:2021vx} and superconducting qubit~\cite{Braumuller:2017wj,Yoshihara:2017ur} setups, where continuous readout is a well-established tool. As a concrete example, 
we outline in the following a possible implementation of our sensing scheme with trapped ions, and defer a microscopic derivation of the LME \eqref{eq:LME_Rabi} in such a setup to Appendix~\ref{appD}.

\begin{figure}[t!]
\centering{} \includegraphics[width=0.49\textwidth]{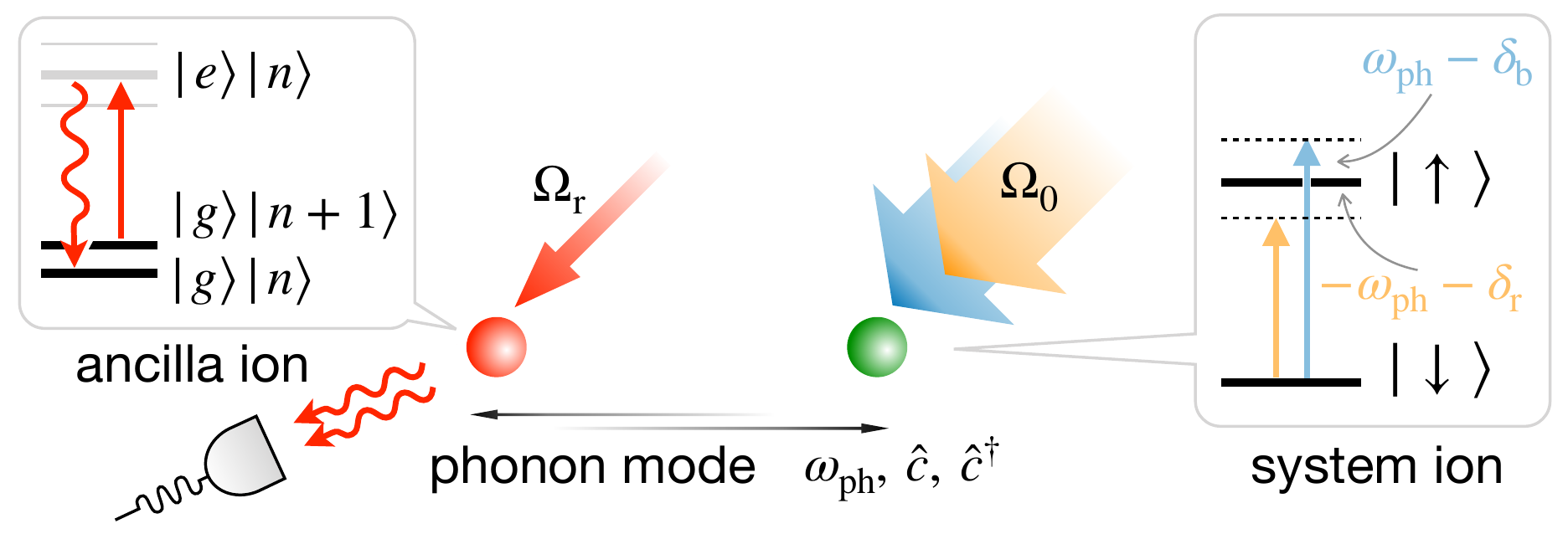} 
\caption{Implementation of our criticality-enhanced sensing scheme in a trapped-ion setup. Two ions in a linear Paul trap share a collective phonon mode. The system ion is driven by bichromatic laser beams tuned close to the blue and red sidebands, realizing the Rabi Hamiltonian \eqref{eq: Hamiltonian}. The ancilla ion is subjected to weak sideband resolved laser cooling, of which the scattered photons are collected and counted continuously by a photon detector, effectively accomplishing continuous counting of the phonons (see the text).}
\label{fig:fig6} 
\end{figure}

As shown in Fig.~\ref{fig:fig6}, we consider two ions trapped in a linear Paul trap sharing a quantized vibrational motion (phonon mode), which we assume is cooled to the ground state. We assume that the two ions can be manipulated independently by lasers, e.g. via single-ion addressability with focussed laser beams~\cite{Linke3305} or via frequency space addressing using a crystal of mixed species~\cite{Negnevitsky:2018ur}. The system ion possesses a pair of internal states $\{\ket{\downarrow},\ket{\uparrow}\}$ representing a (pseudo)spin-$1/2$, which, together with the phonon mode, provides the two elementary degrees of freedom for implementing the Rabi Hamiltonian~\eqref{eq: Hamiltonian}. The ancilla ion, subjected to (weak) sideband resolved laser cooling, introduces controllable dissipation to the phonons and allows for continuous readout of the phonon properties via detection of the fluorescence photons.

Following Refs.~\cite{Pedernales:2015vl,Puebla_2016,PhysRevA.95.063844,PhysRevLett.118.073001,PhysRevX.8.021027,Cai:2021vx}, the Rabi Hamiltonian~\eqref{eq: Hamiltonian} can be engineered by driving the spin transition $\ket{\downarrow}\to\ket{\uparrow}$ with bichromatic laser beams at the same Rabi frequency $\Omega_{0}$, and respectively detuned by $\omega_{\rm ph}-\delta_{\rm b}$ and $-\omega_{\rm ph}-\delta_{\rm r}$ with respect to the ionic resonance; cf.~Fig.~\ref{fig:fig6}. Here $\omega_{\rm ph}$ is the phonon frequency, and $\delta_{\rm b(r)}\ll \omega_{\rm ph}$ are small frequency offsets. As such, the two laser beams are nearly resonant to the blue (red) phonon sideband, thus inducing transition processes $\sim\hat{\sigma}_+ \hat{c}^\dag$ ($\hat{\sigma}_+ \hat{c}$) respectively, with $\hat{\sigma}_+=\ket{\uparrow}\bra{\downarrow}$, $\hat{\sigma}_z=\ket{\uparrow}\bra{\uparrow}-\ket{\downarrow}\bra{\downarrow}$ the Pauli operators and $\hat{c}$ $(\hat{c}^\dag)$ the creation (annihilation) operator of the phonon mode. By moving to a suitable rotation-frame and keeping all nearly-resonant processes~\cite{Pedernales:2015vl}, we arrive at the effective Hamiltonian
\begin{equation}
\label{eq:ion_Rabi}
\hat{H}=\frac{\delta_{\rm b}-\delta_{\rm r}}{2}\hat{c}^\dag\hat{c}+\frac{\delta_{\rm r}+\delta_{\rm b}}{4}\hat{\sigma}_z
+\frac{\eta_{\rm LD}\Omega_0}{2}\hat{\sigma}_x(\hat{c}+\hat{c}^\dag),
\end{equation}
where $\eta_{\rm LD} =k/\sqrt{2m\omega_{\rm ph}}$ is the Lamb-Dicke parameter, with $m$ the ion mass and $k$ the magnitude of the laser wavevector along the direction of the quantized oscillation. Equation \eqref{eq:ion_Rabi} recovers the Rabi Hamiltonian \eqref{eq: Hamiltonian} via the identification $\omega=(\delta_{\rm b}-\delta_{\rm r})/2$, $\Omega=(\delta_{\rm b}+\delta_{\rm r})/2$ and $\lambda=-\eta_{\rm LD}\Omega_0/2$. Importantly, all three parameters can be adjusted independently in experiment, allowing for tuning of the system to the CP.

The demonstration of our critical sensing scheme in this system further requires continuous measurement of the phonons, which can be achieved via the ancilla ion; cf.~Fig.~\ref{fig:fig6}. We assume that the ancilla is driven by a readout laser on a dipole-allowed transition $\ket{g}\to\ket{e}$, and the associated scattered light is collected (e.g., via a lens setup) and counted by a photon detector. The laser is tuned in resonance with the red phonon sideband, $\Delta_{\rm e}=-\omega_{\rm ph}$, under the resolved-sideband condition $\omega_{\rm ph}\gg \Gamma_{\rm e},\Omega_{\rm r}$, where $\Gamma_{\rm e}$ is the spontaneous emission rate of the cooling transition $\ket{e}\to\ket{g}$, and $\Omega_{\rm r}$ and $\Delta_e$ are the Rabi frequency and the detuning of the readout laser respectively. Physically, such a condition prescribes that the emission of a fluorescence photon is accompanied by the annihilation of a vibrational phonon due to energy conservation and as a result, continuous counting of fluorescence photons accomplishes counting of phonons. A quantitative theoretical description can be derived by eliminating adiabatically the ancilla ion and projecting out the environment following a procedure similar to that used in Ref.~\cite{Yang:2020vu}. This results in a stochastic ME for the (unnormalized) conditional density matrix of the joint system including the system-ion and the phonon mode
\begin{align}
\label{eq:sme_eliminate}
d\tilde{\rho}_c =& \left(-i[\hat{H},\tilde{\rho}_c]-\frac{\kappa}{2}\{\hat{c}^\dag\hat{c},\tilde{\rho}_c\}+(1-\epsilon)\kappa\hat{c}\tilde{\rho}_c\hat{c}^\dag\right)dt\\
&+ (\epsilon\kappa dt \hat{c}\tilde{\rho}_c\hat{c}^\dag-\tilde{\rho_{c}})dN(t),\nonumber
\end{align}
where $\kappa \simeq k_0^2\Omega_{\rm r}^2/(4\Gamma_e m_{A}\omega_{\rm ph})$ is the (adjustable) damping rate of the phonon mode with $k_0$ the magnitude of the wavevector of the readout laser and $m_A$ the mass of the ancilla ion. The overall detection efficiency $\epsilon\leq 1$ is dependent on various factors, including the collection efficiency of the lens setup and the detector efficiency. In the ideal limit $\epsilon=1$, Eq.~\eqref{eq:sme_eliminate} recovers our model, Eq.~\eqref{eq:sse_photon_counting}, via the identification $\tilde{\rho}_c=\ket{\tilde{\psi}_c}\bra{\tilde{\psi}_c}$. Remarkably, the scaling behavior of the FI persists even at finite efficiency $\epsilon<1$~\cite{Theodore}, allowing for the demonstration of the criticality-enhanced precision scaling under realistic experimental conditions. 
 
A photon collection efficiency of approximately $15\%$ is feasible via lens setups in trapped-ion experiments~\cite{PhysRevLett.96.043003}. Further improvement can be achieved by coupling the ancilla ion to optical cavities~\cite{doi:10.1063/1.4838696}, by simultaneous detection of the fluorescence of several ancilla, or by entanglement-based amplification~\cite{Hempel:2013tt}.

\section{CONCLUSION AND OUTLOOK}
\label{sec: conclusion}
In contrast to the previous state of the art, we have established a protocol for criticality-enhanced sensing via continuous observation of the radiation quanta emitted by critical open sensors. The resulting precision achieves significantly enhanced scaling, thereby establishing the metrological usefulness of the emitted radiation quanta. To achieve this, we have followed a twofold approach.

First, we establish a scaling theory for the global QFI at continuous dissipative CPs. Under general assumptions, we demonstrate that the global QFI obeys transient and long-time scaling laws governed by the universal critical exponents of the underlying CP. Such scaling can be super-Heisenberg with respect to the interrogation time $t$ or the particle number $N$, but not both. We show that the global QFI is close to and can in principle saturate the HL at dissipative criticality, therefore providing rich opportunities towards criticality-enhanced quantum sensing. To achieve such a precision limit, however, requires the most general measurement of the joint system and environment, which may be challenging in practice.

In view of this, we present a feasible sensing scheme for approaching such criticality-enhanced precision scaling, based on continuous measurement of the radiation quanta emitted by the open sensor. We illustrate our protocol via counting photons emitted by a critical open sensor---the open Rabi model. The relatively simple structure of this model allows us to extract the FI of the detected signal, as a key parameter quantifying the achievable precision of our protocol. Similar to the global QFI, the FI manifests (transient and long-time) scaling behavior at the CP that, importantly, exceeds the SQL and is therefore criticality enhanced. Moreover, as the QFI of the reduced density matrix of the open Rabi model obeys the SQL~\cite{PhysRevLett.124.120504}, our protocol outperforms any protocol based on direct measurement of the open sensor alone. Such a continuous-measurement-based sensing scheme can be applied to various open quantum sensors permitting continuous readout, thereby establishing a general and practical strategy for criticality-enhanced sensing with open quantum sensors. 

The quantum sensing framework established in the present work is timely and feasible in view of the significant experimental progresses in recent years towards the integration of synthetic many-body systems as quantum sensors. Promising candidates for directly implementing our model system include the recent realization of the critical Rabi Hamiltonian in trapped-ion setups~\cite{Cai:2021vx}, where continuous readout is a well-established tool. Several experimental platforms have demonstrated the intimately related open Dicke model~\cite{Baumann:2010wv,Zhiqiang:17}, which lies in the same universality class as the open Rabi model and therefore shares the same criticality-enhanced precision. Other relevant systems where our general scaling theory may apply include optically addressable spins in solid, e.g., color centers in diamond~\cite{PhysRevLett.120.150501}, two-dimensional surfaces on diamond~\cite{Cai:2013wm}, driven-dissipative atomic gas~\cite{Diehl:2011tc,PhysRevA.84.031402}, polariton condensates~\cite{boulier2020microcavity} and many-body cavity QED~\cite{RevModPhys.85.553} and circuit QED~\cite{Ma:2019um} setups. With its broad applicability, our sensing protocol may lead to significant improvements in the design of ultimate sensing devices on the basis of interacting many-body systems that may, for example, find applications in human-machine interfaces~\cite{Zhangeaba8792} or heart diagnostics~\cite{Jensen:2018vt}.

The present work raises a few interesting theoretical questions as well. For example, while the general scaling theory of the global QFI (cf. Sec.~\ref{sec: global_QFI}) established here is validated numerically for the open Rabi(Dicke) universality class, we believe that it and the underlying scaling assumptions hold for a broad class of dissipative CPs, including CPs in higher spatial dimensions. Extending such analysis to models belonging to other universality classes and evaluating the associated metrological precision limit therefore represents an attractive theoretical problem. 
In the broad context of open quantum sensors, remaining open questions include how to achieve the criticality-enhanced scaling of the global QFI via realistic measurement schemes, and the possible generalization of present results to non-Markovian scenarios~\cite{PhysRevLett.100.180402}. Finally, as the QFI is a witness of multipartite entanglement~\cite{PhysRevA.85.022321,PhysRevA.85.022322}, the universal scaling of the global QFI at dissipative criticality may lead to interesting implications, e.g., to the study of the dynamics of entanglement and other correlations in critical open quantum systems.

\section*{Acknowledgments}
We acknowledge Klaus M{\o}lmer for helpful comments on the manuscript. This work is supported by the ERC Synergy grant HyperQ (Grant No. 856432), the EU projects HYPERDIAMOND (Grant No. 667192) and AsteriQs (Grant No. 820394),QuMicro (Grant No. 964818), the QuantERA project NanoSpin (13N14811) and the BMBF project DiaPol (13GW 0281C). We acknowledge support from the state of Baden-Württemberg through bwHPC and the German Research Foundation (DFG) through Grant No INST 40/575-1 FUGG (JUSTUS 2 cluster). Part of the numerical simulations were performed using the QuTiP library~\cite{JOHANSSON20131234}.

\appendix
\section{JUSTIFICATION OF THE DYNAMIC SCALING ANSATZ}
\label{appA}
In this appendix, we discuss in detail the dynamic scaling ansatz ~\eqref{scaling_correlators} underlying the universal scaling of the global QFI, by combining theoretical analysis from the renormalization group perspective and numerical validation for the specific model system demonstrating our sensing scheme. 

\subsection{Theoretical analysis}
Equation~\eqref{scaling_correlators} is a generalization of the established dynamic scaling form given in Eq.~\eqref{eq:S_st_scaling} via the inclusion of the inverse evolution time $\tau^{-1}$ as a new scaling variable. Crucial for such a generalization is the assumption that $\tau^{-1}$ is a relevant scaling variable under the renormalization group (RG) transformation, with a scaling dimension identical to the dynamic critical exponent $z$ of the dissipative CP:
\begin{align}
\label{eq:fixed_point_assumption}
\tau^{-1}\to\lambda^z \tau^{-1}, \,\,\, {\rm under\,the\,RG}\,\,\,L\to \lambda^{-1} L\,,
\end{align}
in which $L$ denotes the system size and $\lambda>1$ is a rescaling factor. Below we show that the assumption ~\eqref{eq:fixed_point_assumption} and, thus Eq.~\eqref{scaling_correlators}, are valid at generic continuous CPs that are not multicritical, i.e., that are not the intersection of two or more critical curves that belong to different universality classes. In the (relatively rare) case that the dissipative CP is multicritical, Eq.~\eqref{scaling_correlators} may need to be replaced by more sophisticated scaling forms reflecting the crossover physics between multiple universality classes.

Let us start our discussion by analyzing an analogous scenario in quantum phase transitions of equilibrium statistical mechanics. Consider an equilibrium system described by the thermal partition function $Z_{\rm T}={\rm tr}[e^{-\beta \hat{H}(B)}]$, with $\beta=1/{k_B T}$ the inverse temperature and $\hat{H}(B)$ the system Hamiltonian dependent on a parameter $B$ (we consider a single relevant parameter for simplicity).
Suppose that at $T=0$ the system undergoes a continuous quantum phase transition at a quantum CP $B_{\rm CP}$, say $B_{\rm CP}=0$. Consequently, thermodynamic quantities obey universal scaling behavior close to $B_{\rm CP}$. As the simplest example, the singular part of the susceptibility of a local observable $\hat{o}$ obeys
\begin{equation}
\label{eq:local_op_scaling}
\partial_B\langle\hat{o}\rangle=\Lambda^{1/\nu-\Delta_{\hat{o}}}\Phi_0 \left(\Lambda  L^{-1},\Lambda ^{1/\nu }B\right),
\end{equation}
where $L$ is the system size, $\Lambda$ is the RG cutoff scale and $\Delta_{\hat{o}}$ and $\nu$ are critical exponents. Equation \eqref{eq:local_op_scaling} can be theoretically constructed within the RG framework. The CP corresponds to a RG fixed point, and $L^{-1}$ and $B$ are the (only) two relevant scaling variables of the RG transformation, which transforms by a simple rescaling as $L^{-1}\to \lambda L^{-1}$ and $B\to \lambda^{1/\nu}B$ close to the fixed point. According to the scaling assumption, all parameters can be expanded in the basis of the scaling variables close to criticality. Expanding various thermodynamic quantities in $L^{-1}$ and $B$ and performing repeated RG transformation leads to the universal scaling forms as exemplified by Eq.~\eqref{eq:local_op_scaling}.

\begin{figure}[t!]
\centering{}
\includegraphics[width=0.48\textwidth]{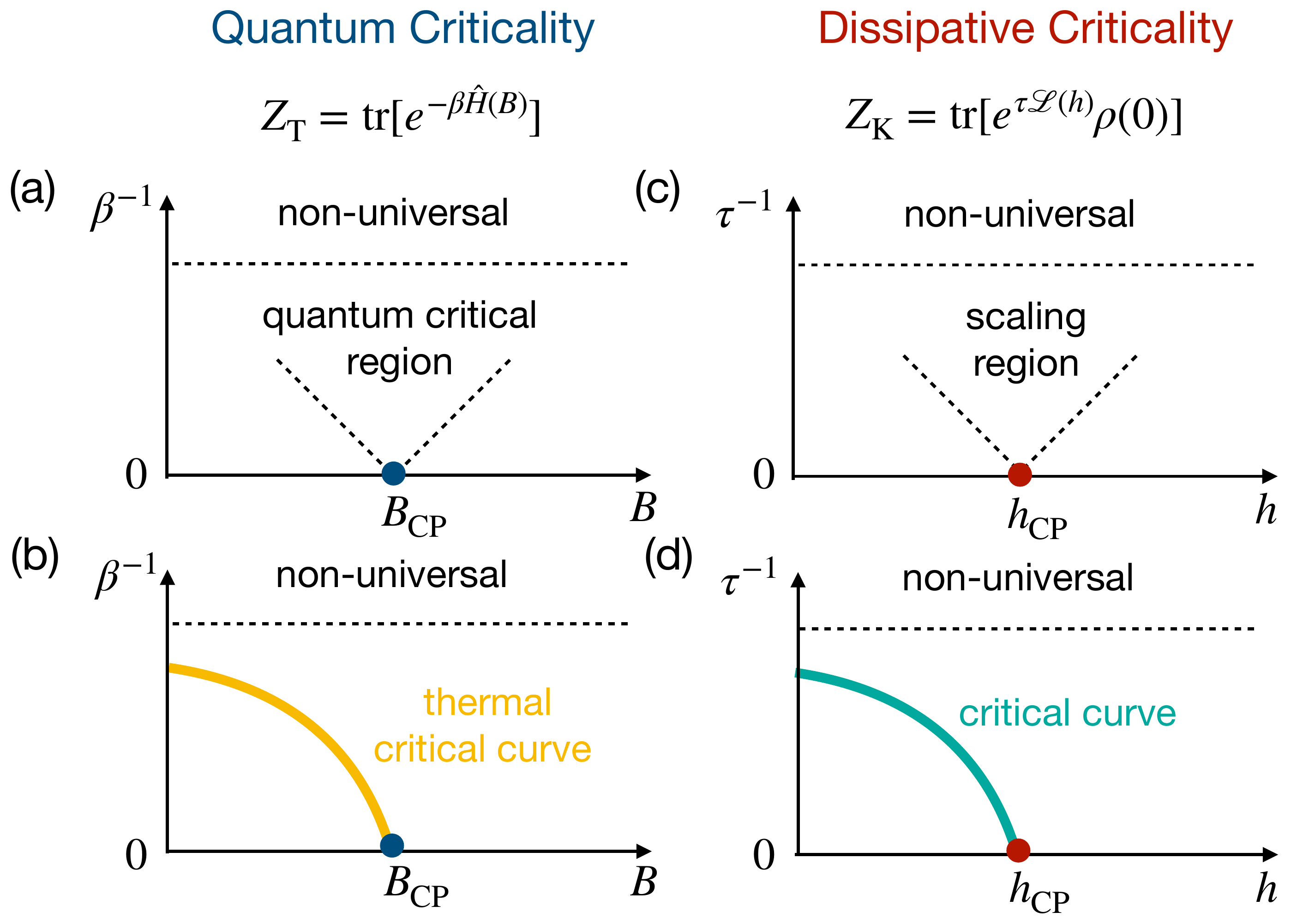} 
\caption{Scenarios of quantum CPs and dissipative CPs. (a) The critical properties of a quantum CP extends to a quantum critical region characterized by a nonzero temperature $\beta^{-1}>0$, which can be captured by scaling laws such as Eq.~\eqref{eq:local_op_scaling_nonzero_T}. (b) A quantum CP as the end point of a thermal critical curve. In this case simple scaling laws such as Eq.~\eqref{eq:local_op_scaling_nonzero_T} may fail to describe the physics at a non-zero temperature $\beta^{-1}>0$. (c)-(d) Analogous situations for dissipative CPs. For dissipative CPs, the scaling variable is not the temperature $\beta^{-1}$, but the inverse evolution time $\tau^{-1}$.}
\label{fig:fig9} 
\end{figure}

Now consider the system behavior at a slightly nonzero temperature $T>0$, inside the celebrated ``quantum critical region''; cf. Fig.~\ref{fig:fig9}(a). This region is characterized by a RG flow from the unstable fixed point at $T=0$ (corresponding to the quantum CP) to the high-$T$ stable fixed point (corresponding to the disordered phase). In this region, the temperature $T$ enters as a new relevant scaling variable besides $L^{-1}$ and $B$. Its scaling dimension can be read by a simple dimensional analysis of the partition function $Z_{\rm T}$---as $ \beta$ has the dimension of (imaginary) time $t$, and $t$ transforms as $t\to \lambda^{-z} t$ under the RG, with $z$ the dynamic critical exponent of the quantum CP. Therefore,
\begin{equation}
\label{eq:fixed_point_QPT}
\beta^{-1}\to\lambda^z \beta^{-1}, \,\,\, {\rm under\,the\,RG}\,\,\,L\to \lambda^{-1} L.
\end{equation}
In the quantum critical region, we can again expand all thermodynamic quantities in the basis of the scaling operators $L, B$, and $\beta^{-1}$, and perform repeated RG transformation, therefore extending the zero-temperature universal scaling laws to $T>0$. As it turns out, this amounts to simply adding $\Lambda^z \beta^{-1}$ as a new variable in the scaling function $\Phi$ (which is different from the zero-temperature scaling function $\Phi_0$). For example, Eq.~\eqref{eq:local_op_scaling} is extended to
\begin{equation}
\label{eq:local_op_scaling_nonzero_T}
\partial_B\langle\hat{o}\rangle=\Lambda^{1/\nu-\Delta_{\hat{o}}}\Phi \left(\Lambda  L^{-1},\Lambda ^{1/\nu }B,\Lambda^z \beta^{-1}\right),
\end{equation}
and similarly for other thermodynamic quantities, e.g., correlators and response functions. In this way, the quantum criticality extends from the CP to $T>0$---a generic feature of quantum phase transitions. By further increasing $T$ beyond some cutoff temperature (whose exact value depends on the microscopic details of the model), however, these universal behaviors are lost and the system enters a nonuniversal region, as demonstrated in Fig.~\ref{fig:fig9}(a).

The above paradigm is applicable to describe generic continuous quantum CPs, but not all. The exceptions are multicritical quantum CPs. A typical example is shown in Fig.~\ref{fig:fig9}(b), where the quantum CP, $(B,T)=(B_{\rm CP},0)$, is the end point of a continuous thermal phase transition line that belongs to a different universality class from the quantum CP. As such, the physics in the low-$T$ region is determined not only by the quantum CP but also by the thermal CPs, therefore typically demonstrating more sophisticated crossover behavior than that captured by Eq.~\eqref{eq:local_op_scaling_nonzero_T}. In particular, the dynamic critical exponent $z|_{T=0^+}$ associated with the thermal CP $(B,T)=(B_{\rm CP},T=0^+)$ infinitesimally above the quantum CP can be different from the exponent of the quantum CP, $z|_{T=0^+}\neq z$. From the RG point of view, this reflects that the RG transformation is nonanalytic at a multicritical quantum CP, rendering the extension from Eq.~\eqref{eq:local_op_scaling} to Eq.~\eqref{eq:local_op_scaling_nonzero_T} presented above invalid. Methods to construct appropriate scaling forms to describe the crossover physics near multi-critical CPs are established (see, e.g., Chap. 10-14 of Ref.~\cite{sachdev_2011} for an introduction), but are irrelevant to our analysis here. 

We now turn to the universal physics near a dissipative CP that, as we show below, shares strong similarities with the scenario discussed above. In contrast to equilibrium systems, which are described by a thermal partition function $Z_{\rm T}={\rm tr}[e^{-\beta \hat{H}}]$, (Markovian) driven dissipative systems are described by a Keldysh partition function $Z_{\rm K}={\rm tr}[\rho(\tau)]={\rm tr}[e^{\tau{\cal L}}\rho(0)]$ (such a description is completely equivalent to the Lindblad-master-equation description, we refer to Ref.~\cite{Sieberer_2016} for an introduction of the Keldysh framework). In $Z_{\rm K}$ the evolution time $\tau$ plays a role analogous to $\beta$ in $Z_{\rm T}$. Analogous to a quantum CP, which is defined by the non-analyticity of $Z_{\rm T}$ on the $\beta=\infty$ (i.e., $T=0$) plane, a dissipative CP is characterized by the non-analyticity of $Z_{\rm K}$ on the $\tau=\infty$ plane, i.e., of the steady state. There is an important difference, however---as $\hat{H}$ is the Hermitian, $Z_{\rm T}$ by construction obeys the detailed balance condition (as reflected by the quantum fluctuation-dissipation theorem), such feature is lost in $Z_{\rm K}$.

Crucially, our analysis of the quantum CP presented above does not rely on the assumption of thermal equilibrium (i.e., the condition of detailed balance). Its validity essentially relies on two conditions: (1) the parameter $\beta$ enters as a linear coefficient in ${\rm log}(Z_{\rm T})$, such that $\beta$ has the dimension of time; (2) the quantum CP is not multicritical, such that the RG transformation is differentiable at the quantum CP. These conditions are also fulfilled in the case of a dissipative CP (we note that, although the initial state $\rho(0)$ formally enters the definition of $Z_{\rm K}$, it contributes only irrelevant parameters under the RG as long as the dissipative system has a unique steady state~\cite{Sieberer_2016}). Therefore, all the RG reasoning presented above can be adapted to the case of dissipative CP by simply replacing $\beta$ with $\tau$. Specifically we have two conclusions.

(I) The critical properties of a generic dissipative CP (defined for $\tau=\infty$) extends to a scaling region characterized by long, but finite evolution time $\tau$; cf., Fig.~\ref{fig:fig9}(c). Such a region terminates at a short time scale $t_{\rm cutoff}$, whose exact value depends on the model details, and the evolution shorter than $t_{\rm cutoff}$  is nonuniversal. In the scaling region, if the dissipative CP is not multicritical, the inverse evolution time $\tau^{-1}$ plays the role of a relevant scaling variable, with scaling dimension identical to the dynamic critical exponent $z$ of the dissipative CP:
\begin{equation}
\label{eq:fixed_point_justification}
\tau^{-1}\to\lambda^z \tau^{-1}, \,\,\, {\rm under\,the\,RG}\,\,\,L\to \lambda^{-1} L.
\end{equation}
This justifies our assumption ~\eqref{eq:fixed_point_assumption}. Hence, universal scaling laws at $\tau=\infty$ can be extended to finite $\tau$ by simply adding $\Lambda^z \tau^{-1}$ as a new variable in the scaling function. Applying this to the steady-state dynamic structure [cf. Eq.~\eqref{eq:S_st_scaling}], we arrive at our ansatz ~\eqref{scaling_correlators}. 

(II) In the rare case that the dissipative CP is multicritical, assumption ~\eqref{eq:fixed_point_assumption} is not necessarily valid, and may need to be replaced by more sophisticated scaling forms to reflect the underlying crossover physics. While multicritical dissipative CPs are to a large extent unexplored, we conjecture a possible crossover scenario in Fig.~\ref{fig:fig9}(d) via analogy to the equilibrium situation. A detailed investigation of the universal physics of multicritical dissipative CPs is attractive, is nevertheless beyond the scope of our paper.

\subsection{Numerical validation for the open Rabi model}
Here we provide a numerical validation of the dynamic scaling ansatz \eqref{scaling_correlators} for the model system that demonstrates our sensing scheme. Equation~\eqref{scaling_correlators} prescribes that the correlator $S(\tau,s,L)$ obeys a universal scaling form determined by the three relevant perturbations $h$, $\tau^{-1}$, and $L^{-1}$, which can be validated via finite-size scaling collapse in the $(h,\tau^{-1}, L^{-1},s^{-1})$ parameter space. For illustration, in the following we fix either $\tau^{-1}$ or $L^{-1}$, and show numerical finite-size scaling of the correlator with respect to the other two perturbations. To be specific, we choose $\hat{O}=\hat{n}$, which has scaling dimension $\Delta_{\hat{n}}=-1/2$ (cf.~Sec.~\ref{sec:Quantum_Rabi}).

\begin{figure}[t!]
\centering{} \includegraphics[width=0.49\textwidth]{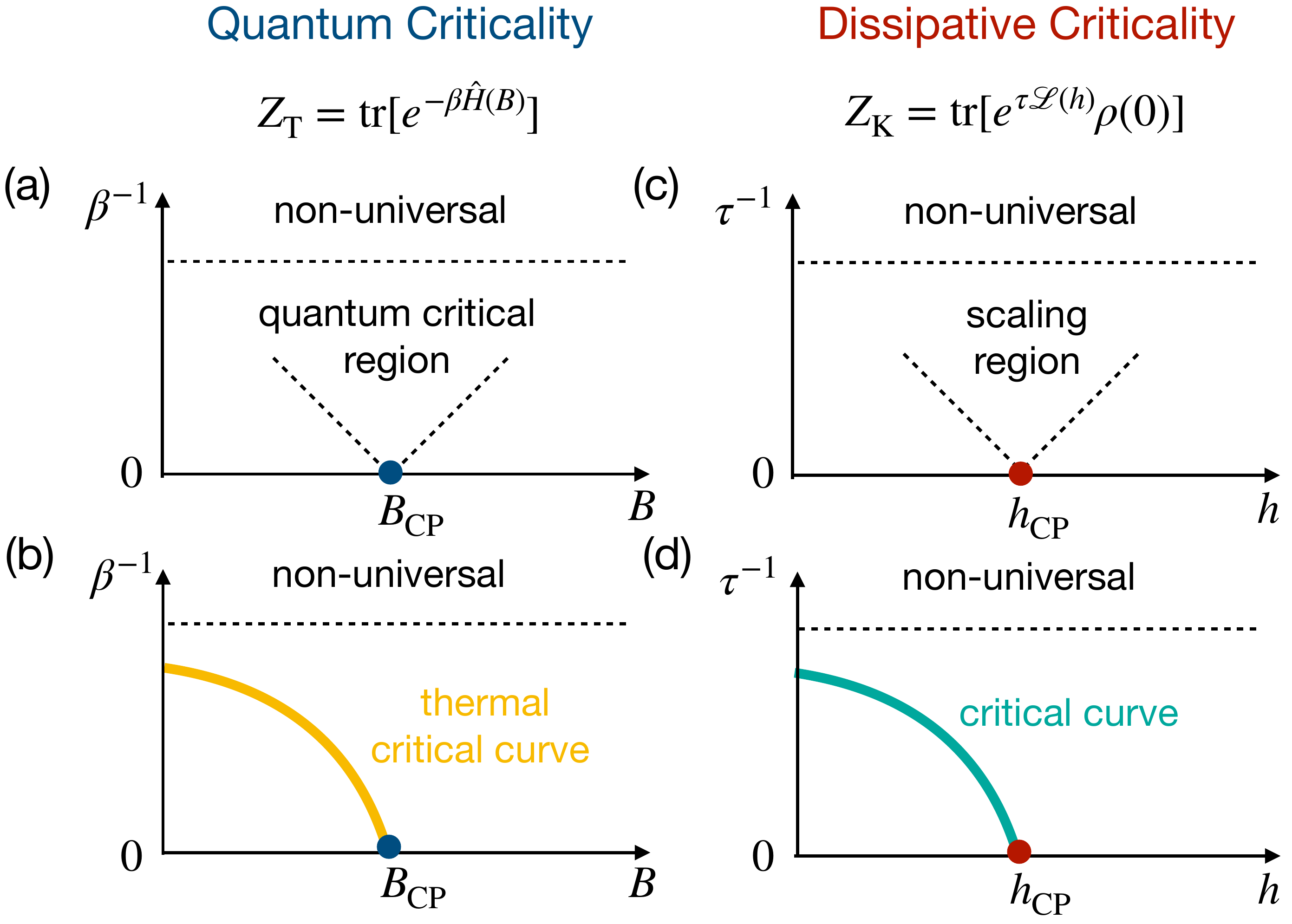} 
\caption{Numerical verification of the dynamic scaling ansatz \eqref{scaling_correlators} for the open Rabi model, which has $z=1$, $\nu=2$ and $\Delta_{\hat{n}}=-1/2$. The perfect scaling collapse demonstrates the validity of Eq.~\eqref{scaling_correlators} along the specific axis in the parameter space (see the text).}
\label{fig:fig_correlators} 
\end{figure}

First, we fix $\tau^{-1}=0$, which corresponds to taking the infinitely long time limit, at which $S(\tau,s,\eta)=S_{\rm st}(s,\eta)$. In Fig.~\ref{fig:fig_correlators} (a), we show the results of our numerical finite-size scaling of $S_{\rm st}(s,\eta)$ for different $\eta$ and $h=g-g_{\rm CP}$ at various $\kappa s$ (which serves as an irrelevant perturbation). The perfect scaling collapse indicates that, at sufficiently large $|h|$, $|h|\gg \eta^{-1/\nu}$, $h$ is the dominant perturbation and, correspondingly $S_{\rm st}(s,\eta)\sim  |h|^{2\nu\Delta_{\hat{n}}}=h^{-2}$. In contrast, at small $h$ the inverse system size $\eta^{-1}$ is the dominant perturbation, and, correspondingly $S_{\rm st}(s,\eta)\sim \eta^{-2\Delta_{\hat{n}}}=\eta$. These demonstrate the validity of Eq.~\eqref{scaling_correlators} along the axis $\tau^{-1}=0$.

Next, we fix $h=0$, and perform finite-size scaling of $S(\tau,s,\eta)$ for different $\eta$ and $\tau$ at various $\kappa s$, as shown in Fig.~\ref{fig:fig_correlators} (b). The perfect scaling collapse indicates that, at sufficiently small $\tau^{-1}$,  the inverse system size $\eta^{-1}$ is the dominant perturbation and, correspondingly $S(\tau,s,\eta)\sim \eta^{-2\Delta_{\hat{n}}}=\eta$, whereas at large $\tau^{-1}$ the scaling form becomes $S_{\rm st}(s,\eta)\sim  \tau^{-2\Delta_{\hat{n}}/z}=\tau$. These demonstrate the validity of Eq.~\eqref{scaling_correlators} along the axis $h=0$.

A similar procedure can be carried out for all the parameter space $(h,\tau^{-1}, L^{-1},s^{-1})$ surrounding the CP, which validates Eq.~\eqref{scaling_correlators}.

\section{NUMERICAL CALCULATION OF THE GLOBAL QUANTUM FISHER INFORMATION}
\label{appB}
We follow an efficient method proposed in Ref.~\cite{PhysRevLett.112.170401} to numerically calculate the global QFI of the joint system and environment, which we summarize here to keep our work self-contained. The starting point is an alternative expression of the QFI equivalent to Eq.~\eqref{eq: gQFI_1} of the main text,
\begin{equation} \label{QFI_equivalent}
I_\theta (t)= 4 \partial_{\theta_1} \partial_{\theta_2} (\log |\bra{\Psi_{\theta_1}(t)} \Psi_{\theta_2}(t) \rangle|)|_{\theta_1=\theta_2=\theta},
\end{equation}
where $|\Psi_{\theta}(t)\rangle$ is the $\theta$-dependent pure global state of the system plus the environment at time $t$.
Although a complete knowledge of the global quantum state is impractical as the number of photons in the environment increases with time, in Ref.~\cite{PhysRevLett.112.170401} the authors describe an efficient way to calculate the QFI without accessing the full quantum state.  The main idea is that as long as the Born-Markov approximation is valid, we can descretize time such that in every time interval $[t_i,t_i+d t)$ the system interacts with independent environmental degrees of freedom. As a result, at the final time $t=n d t$, the state of the joint system and environment $\ket{\Psi(t)}$ can be expressed as an entangled state in the tensor-product Hilbert space of the open system and $N$ environmental subspaces:
\begin{equation} \label{eq:evolution_total_system}
\ket{\Psi(t)}=\hat{U}_{t_{n-1}} \hat{U}_{t_{n-2}} \cdots \hat{U}_{t_0} \ket{\psi(0)}\ket{0_{n-1},\cdots 0_0},
\end{equation}
whith $|\psi(0)\rangle$ the initial state of the system and $\hat{U}_{t_i}$ the unitary operators acting on the joint system and environment state associated with the time bin $[t_i,t_i+dt)$. Here we have also associated an $l$-dimensional Hilbert space for each environment subspace $\{\ket{m_i}\},m=0,1,\dots,l-1$, and assumed that each of them are initialized in the vacuum $\ket{0_i}$ before the interaction. In turn, we can introduce ``measurement effect operators'' $\hat{M}_{m_i}$ that define the evolution of the open quantum system with the associated transfer of the environment from state $\ket{0_i}$ to $\ket{m_i}$. This allows us to write 
\begin{align}
 \label{eq_ef_measurement_operatos}
\ket{\Psi(t)}=\sum_{m_0 \cdots m_{n-1}=0}^{l-1}& \hat{M}_{m_{n-1}} \cdots \hat{M}_{m_0}
 \ket{\psi(0)}\nonumber\\
&\otimes \ket{m_{n-1},\cdots m_0}
\end{align}

A particular choice of $m_0 \cdots m_{n-1}$ that corresponds to a single term in Eq. (\ref{eq_ef_measurement_operatos}) defines the quantum stochastic trajectory $\ket{\psi_c (t)} \propto \hat{M}_{m_{n-1}} \cdots \hat{M}_{m_0} \ket{\psi(0)}$. The unknown parameter, $\theta$, that we want to estimate is encoded in the unitary operators $\hat{U}_{t_i}$ and hence the measurement effect operators  $\hat{M}_{m_i}$. After tracing out the environment degrees of freedom we obtain the reduced density matrix that describes the open quantum system and evolves according to the equation
\begin{equation} \label{eq:evolution_reduced_dm}
\frac{d\rho(t)}{dt}=\frac{1}{dt}\left(\sum_{m=0}^{l-1} \hat{M}_{m} \rho(t) \hat{M}_{m}^{\dagger}-\rho(t)\right).
\end{equation}
For infinitesimal time step $dt$, this gives rise to the standard LME as described in the main text.
Going back to Eq.~(\ref{QFI_equivalent}), we note that the inner product $\bra{ \Psi_{\theta_1}} \Psi_{\theta_2} \rangle$ can be written as ${\rm Tr}_{\rm sys,env} \{ \ket{\Psi_{\theta_1}} \bra{\Psi_{\theta_2}} \}={\rm Tr}_{\rm sys} \{\rho_{\theta_1,\theta_2} \} $, where the action of the operators $\hat{M}_{m_{n-1}}(\theta_1) \cdots \hat{M}_{m_0}(\theta_1) $ from the left and $\hat{M}_{m_0}^{\dagger}(\theta_2) \cdots \hat{M}_{m_{n-1}}^{\dagger}(\theta_2)$ from the right has been absorbed in the definition of $\rho_{\theta_1,\theta_2}$. This is similar to Eq. (\ref{eq:evolution_reduced_dm}) and thus $\rho_{\theta_1,\theta_2}$  is the solution of a generalized ME ${d \rho}/{dt}= \mathcal{L}_{\theta_1,\theta_2} \rho$.
Specifically to our case that the unknown parameter is encoded in the Hamiltonian of the open system, the generalized ME can be derived to be
\begin{equation}
\mathcal{L}_{\theta_1,\theta_2} \rho= -i \hat{H}({\theta_1} )\rho +i \rho \hat{H}({\theta_2})+\sum _\ell \left( \hat{J}_\ell \rho \hat{J}_\ell^\dag 
-\frac{1}{2}\{\hat{J}_\ell^\dag\hat{J}_\ell, \rho\} \right).
\nonumber
\end{equation}
Solving the above equation numerically for $(\theta_1,\theta_2)$ in the neighborhood of $(\theta,\theta)$ allows us to determine the global QFI by numerical difference via Eq.~\eqref{QFI_equivalent}.

\section{SCALING OF THE GLOBAL QUANTUM FISHER INFORMATION AWAY FROM CRITICALITY}
\label{appC}
\begin{figure}[t!]
\centering{} \includegraphics[width=0.49\textwidth]{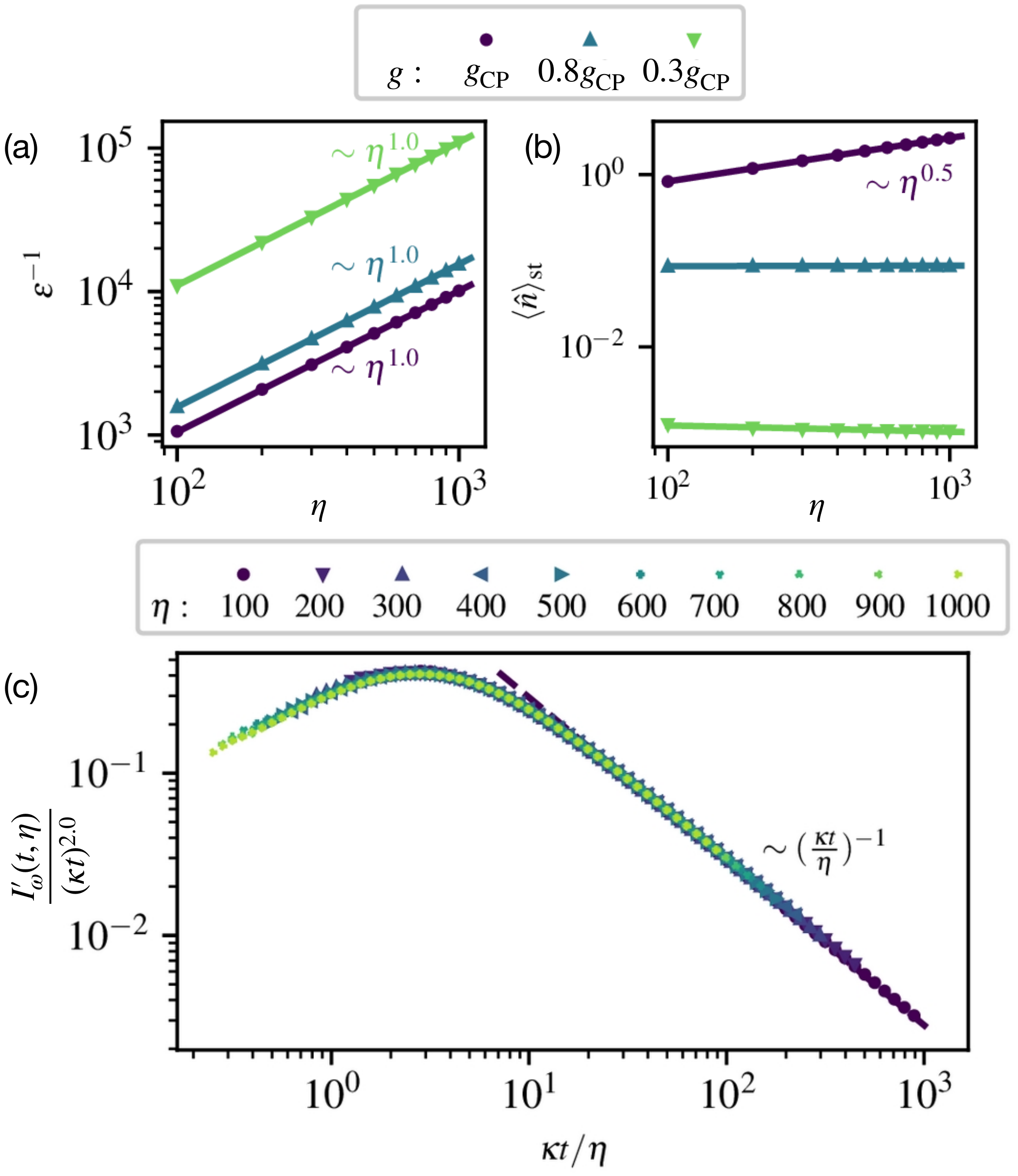} 
\caption{Numerical finite-size scaling of (a) the inverse of the Liouvillian gap $\varepsilon^{-1}$ and (b) the steady mean cavity occupation $\langle\hat{n}\rangle_{\rm st}$ for different system sizes $\eta$ and for different values of $g$. (c) Scaling of the global QFI $I_\omega^\prime(t,\eta)$ when $g$ is tuned away from the CP, $g=0.8g_{\rm CP}$. Parameters: $\kappa=0.1$, $\omega=1$.  }
\label{fig:QFI_away_criticality} 
\end{figure}
In this section we study the behavior of the global QFI of the open Rabi model, assuming that the model is tuned far away from the CP, and contrast it with the universal scaling behavior at the CP as analyzed in the main text. The latter is given by Eqs.~\eqref{eq:QFI_short_time_general} and ~\eqref{eq:QFI_long_time_general} in terms of the spatial dimension $d=0$, the dynamic critical exponent $z=1$, and the scaling dimension $\Delta_{\hat{n}}=-1/2$ (we choose $\hat{o}=\hat{n}$ for illustration). The dynamic critical exponent $z$ characterizes the closing of the Liouvillian gap $\varepsilon$ at the CP with increasing system size via $\varepsilon\sim \eta^{-z}$. The scaling dimension $\Delta_{\hat{n}}$ characterizes the divergence of $\langle\hat{n}\rangle_{\rm st}$ at the CP with increasing system size via $\langle\hat{n}\rangle_{\rm st} \sim {\eta^{-\Delta_{\hat{n}}}}$. Typically, such universal scaling behavior is lost if the system is tuned sufficiently far away from the CP, outside the so-called critical region.

Interestingly, as a zero-dimensional, finite-component model, the open Rabi model possesses unique features regarding the extension of the critical region. Specifically, the dynamic scaling behavior $\varepsilon\sim \eta^{-z}$ extends from the CP to arbitrary $g\neq 0$. This is illustrated in Fig.~\ref{fig:QFI_away_criticality}(a), where we plot the finite-size scaling of $\varepsilon^{-1}$ both at the CP ($g=g_{\rm CP}$) and far away from the CP ($g=0.8g_{\rm CP}$ and $g=0.3g_{\rm CP}$), which all manifest the same scaling $\varepsilon\sim \eta^{-z}$. In contrast, the static scaling $\langle\hat{n}\rangle_{\rm st}\sim {\eta^{-\Delta_{\hat{n}}}}$ manifests only in the critical region $|g-g_{\rm CP}|\lesssim \eta^{-1/\nu}$ with $\nu=2$ the correlation length critical exponent as defined in Sec.~\ref{sec:Quantum_Rabi} of the main text. The loss of the static criticality outside this region is demonstrated in Fig.~\ref{fig:QFI_away_criticality}(b).

Consequently, the global QFI manifests interesting scaling behavior outside the critical region. This is illustrated in Fig. \ref{fig:QFI_away_criticality}(c) with the example of $g=0.8g_{\rm CP}$. In particular, we can still identify the transient regime $\kappa t\lesssim \eta$ and the long-time regime $\kappa t\gg \eta$ in such scaling, as a direct consequence of the persistence of the dynamic criticality. Specifically, Fig.~\ref{fig:QFI_away_criticality}(c) indicates
\begin{numcases}{{I}_\omega^\prime(t,\eta)=}
(\kappa t)^2  {f_I^\prime}\left(\kappa t/\eta\right),&$\kappa t\lesssim \eta$, \label{eq:QFI_short_time_away_criticality}\\
{ {\rm const.}} \times\kappa t \eta,&$\kappa t \gg \eta$, \label{eq:QFI_long_time_away_criticality}
\end{numcases}
where the prime notation in ${I}_\omega^\prime(t,\eta)$ indicates that the global QFI is evaluated away from the CP. We note that these scalings are significantly lower than the scaling at the CP; cf. Eq.~\eqref{eq:QFI_short_time} and \eqref{eq:QFI_long_time}. This is due to the lack of (static) criticality away from the CP.

The scaling laws \eqref{eq:QFI_short_time_away_criticality} and \eqref{eq:QFI_long_time_away_criticality} can be understood by adopting the physical picture developed in Sec.~\ref{sec:physical_discussion_FI} in terms of the physical resources involved. In particular, the long-time scaling \eqref{eq:QFI_long_time_away_criticality} can be expressed as $I_\omega^\prime(t,\eta)\sim \kappa^2 tt_*$ [we note that $\langle\hat{n}\rangle_{\rm st}\sim O(1)$ far away from the CP]. Moreover, the transient scaling \eqref{eq:QFI_short_time_away_criticality} can be recovered from the long-time scaling \eqref{eq:QFI_long_time_away_criticality} by the replacement $\eta\to (\kappa t)^{1/z}$, reflecting the underlying dynamic criticality that persists even away from the CP.

\section{MICROSCOPIC DERIVATION OF THE LINDBLAD MASTER EQUATION FOR THE OPEN RABI MODEL}
\label{appD}

In this section we discuss the validity of LME (\ref{eq:LME_Rabi}) of the main text.
Although Eq.~(\ref{eq:LME_Rabi}) has been extensively used in the literature \cite{PhysRevA.97.013825,PhysRevLett.124.120504}, a rigorous derivation of it based on the microscopic Hamiltonians of the system and the environment remains an outstanding open question. Indeed, Eq. (\ref{eq:LME_Rabi}) is typically regarded as a “phenomenological” LME \cite{2010, 2007, konopik2020local} that is microscopically justifiable only in the weak coupling regime $\lambda \ll \omega, \Omega$ of the Rabi Hamiltonian (\ref{eq: Hamiltonian}), where the cavity mode and the qubit weakly hybridize. Away from this regime, however, the strongly coupled cavity-qubit system interacts with the environment as a whole, and thus the simple form of the Lindblad dissipator involving only the cavity mode [cf. the second term of Eq.(\ref{eq:LME_Rabi})] is not necessarily valid. Our goal here is to provide a microscopic derivation of Eq. (\ref{eq:LME_Rabi}) in the thermodynamic limit $\eta \rightarrow \infty$ (correspondingly, $\omega \ll \lambda \ll \Omega$), based on the experimental implementation we proposeD in Sec.~\ref{sec:Experimental_implementation}. 

As a summary, such a microscopic derivation is possible for two main reasons. First, in the thermodynamic limit, the eigenstates of the Rabi Hamiltonian factorize, i.e., it can be written as a tensor product of the states of the qubit and the cavity. Correspondingly, the Lindblad dissipators involve operators acting only on the cavity mode. The specific form of these eigenstates (corresponding to squeezed cavity state; see below) leads to jump operators that can only be linear combinations of the annihilation and the raising operator of the cavity mode. Second,
in our proposed trapped-ion implementation (cf., Sec.~\ref{sec:Experimental_implementation} of our manuscript), the fast rotating interaction between the cavity-qubit system and the environment ensures that only the cavity annihilation operator is present in the jump operator after the secular approximation. Below we present a fully rigorous derivation of LME~\eqref{eq:LME_Rabi} that develops the two key points summarized above. 

\subsection{System-bath coupling and the Born-Markov approximation}
We start from the total Hamiltonian of the cavity-qubit system and the environment as realized in the trapped-ion setup we proposed in Sec.~\ref{sec:Experimental_implementation},
\begin{equation} \label{total_ham}
\hat{H}_{\rm tot}=\hat{H}+\hat{H}_{ \rm E}+\hat{H}_{\rm int},
\end{equation}
where $\hat{H}$ is the Rabi Hamiltonian (\ref{eq: Hamiltonian}) of the cavity-qubit system, $\hat{H}_{\rm E}$ is the free Hamiltonian of the environment, and $\hat{H}_{ \rm int}$ is the weak system and environment coupling. We assume that $\hat{H}_{ \rm int}$ picks the form
\begin{equation} \label{eq:interact_ham}
\hat{H}_{ \rm int}=\hat{A}(t)\otimes \hat{B}=e^{i \nu_0t \hat{c}^{\dagger} \hat{c}}(\ha+\hadag) e^{-i \nu_0t \hat{c}^{\dagger} \hat{c}} \otimes \hat{B}
\end{equation}
Here $\hat{c}$ $(\hat{c}^{\dagger})$ is the cavity annihilation (creation) operator, $\hat{B}$ is a bath operator to be specified later in Appendices \ref{sec:thermal_bath} and \ref{sec:ancilla_ion}, and $\nu_0$ is a frequency offset that we assume is much larger than the spectral gap of $\hat{H}$. We emphasize that Eq.~\eqref{eq:interact_ham} is time dependent, in contrast to conventional time-independent system-bath coupling. Such fast rotating coupling term naturally emerges in our proposed trapped ion setup, since the Rabi Hamiltonian  is derived in a rotating frame with respect to $\hat{H}_0=\nu_0 \hat{c}^{\dagger} \hat{c}+\frac{1}{2} [\omega_{\rm I}-\frac{1}{2}(\delta_{\rm b}+\delta_{\rm r})]\hat{\sigma}_z$, where $\nu_0=\omega_{\rm ph}-\frac{1}{2}(\delta_{\rm _b}-\delta_{\rm r})$ is a shift from the phonon frequency $\omega_{\rm ph}$ and $\omega_{\rm I}$ is the transition frequency of the ion \cite{Pedernales:2015vl,PhysRevLett.118.073001}. In such a rotating frame, the phonon-environment coupling adopts the form of Eq.~\eqref{eq:interact_ham}, i.e., becomes time dependent. In contrast to $\hat{H}_{ \rm int}$, the formal derivation of the LME is largely independent of the details of $\hat{H}_{\rm E}$ and we leave it unspecified until Appendices \ref{sec:thermal_bath} and \ref{sec:ancilla_ion} when we consider specific cases related to our experimental implementation.

Defining the total system and environment state as $\rho_{\rm tot}$ and moving to the interaction picture via $\tilde{\rho}_{\rm tot}(t)= e^{i(\hat{H}+\hat{H}_{\rm E})t} \rho_{\rm tot} (t) e^{-i(\hat{H}+\hat{H}_{\rm E})t}$, the evolution of the system-environment obeys $\dot{\tilde{\rho}}_{\rm tot}(t)=-i [\tilde{H}_{\rm int},\tilde{\rho}_{\rm tot}(t)]$,
where $\tilde{H}_{\rm int}(t)=e^{i(\hat{H}+\hat{H}_{\rm E})t} \hat{H}_{ \rm int} e^{-i(\hat{H}+\hat{H}_{\rm E})t}$. Following the standard procedure \cite{1998aibp,breuer2002theory,Rivas_2012}, by tracing out the environment we obtain a differential equation for the state of the system alone $\tilde{\rho}(t)={\rm tr_{E}} \{ \tilde{\rho}_{ \rm tot}  (t)\}  \equiv e^{i\hat{H}t} \rho(t) e^{-i\hat{H}t}$
\begin{equation} \label{eq:total_evolution2}
\dot{\tilde{\rho}}(t)=-\int_0^t  {\rm tr_{\rm E}} (\left\{ \left[ \tilde{H}_{\rm int}(t),[\tilde{H}_{\rm int}(s),\tilde{\rho}_{\rm tot}(s)]\right] \right\}) ds,
\end{equation}
where we assume that ${\rm tr_{\rm E}} \{ \tilde{H}_{\rm int}(t) \tilde{\rho}_{\rm tot}(0) \}=0$.
To proceed, we make the Born approximation by assuming that the environment is not significantly affected by the interaction and thus the total state can be factorized as $\tilde{\rho}_{\rm tot}(t) \approx \tilde{\rho}(t) \otimes \rho_{\rm E} $ where $\rho_{\rm E}$ is the state of environment. Further introducing the Markov approximation via replacing $\tilde{\rho}_{\rm tot}(s)$ in Eq.(\ref{eq:total_evolution2}) by $\tilde{\rho}_{\rm tot}(t)$, and extending the lower limit to $-\infty$ and substituting $s\rightarrow t-s$, we deduce a fully Markovian master equation
\begin{equation} \label{eq:markovian_me}
\dot{\tilde{\rho}}(t)=-\int_0^\infty {\rm tr_{\rm E}} (\left\{ \left[ \tilde{H}_{\rm int}(t),[\tilde{H}_{\rm int}(t-s),\tilde{\rho}(t) \otimes  \rho_{\rm E}] \right] \right\}) ds
\end{equation}
Using Eq. (\ref{eq:interact_ham}), the equation above is transformed to
\begin{equation} \label{eq:me_interact}
\begin{aligned}
\dot{\tilde{\rho}}(t)=&\int_0 ^{\infty}  J(s) \left[ -\tilde{A}(t) \tilde{A} (t-s) \tilde{\rho}(t)+\tilde{A}(t-s)\tilde{\rho}(t)\tilde{A}(t) \right] ds\\
&+J(-s) \left[ \tilde{A}(t) \tilde{\rho}(t) \tilde{A}(t-s)- \tilde{\rho}(t) \tilde{A}(t-s) \tilde{A}(t) \right]ds, 
\end{aligned}
\end{equation}
where $J(s)={\rm tr_{\rm E}}  \{\hat{B}(t) \hat{B}(t-s) \rho_{\rm E} \}={\rm tr_{E}}  \{\hat{B}(s)\hat{B}\rho_{\rm_{ \rm E}} \}$ is the bath correlation function and $\tilde{A}(t)= e^{i\hat{H}t} \hat{A}(t) e^{-i \hat{H} t}$ with $\hat{A}(t)$ defined in Eq.~(\ref{eq:interact_ham}). The integral over $s$ can be expressed in terms of the spectrum density of the environment $\Gamma(\omega)=\int_{-\infty}^{\infty} J(s)e^{i\omega s}ds$ as 
\begin{equation}
\int_0^\infty J(s) e^{i \omega s}ds=\frac{1}{2} \Gamma(\omega)+i \lambda_{\rm LS},
\end{equation}
where $\lambda_{\rm LS}$ denotes the Lamb shift that can be absorbed via redefinition of the frequency of the cavity mode and from now on we neglect.

\subsection{Expansion of the master equation in the eigenbasis of the Rabi Hamiltonian} \label{sec:expansion_eigenstates}
The next step of our microscopic derivation is to express the operator $\hat{A}$ in the eigenbasis of the Rabi Hamiltonian $\hat{H}$. As shown in Ref.~\cite{PhysRevLett.115.180404}, in the thermodynamic limit $\eta\to\infty$, $\hat{H}$ can be diagonalized exactly and the eigenstates adopt different forms in the two regimes $g<1$ and $g \geq 1$, where $g=2\lambda / \sqrt{\Omega \omega}$ is the dimensionless coupling constant between the cavity mode and the qubit. To be specific, in the following we consider the case $g<1$. In the opposite case $g \geq 1$, the eigenspectrum of $\hat{H}$ is twofold degenerate, with the corresponding eigenstates belonging to the even and odd sectors of the parity symmetry respectively~\cite{PhysRevLett.115.180404}. Nevertheless, the system and environment coupling Hamiltonian \eqref{eq:interact_ham} has vanishing matrix elements between the two sectors. Consequently, the derivation of the LME  can be done in each parity sector, which follows exactly the same steps as for $g<1$ and eventually results in the same LME.

Focusing on the case $g <1$, the eigenstates can be written in a compact form as $\ket{E_{\alpha,n}}=S[r_{\alpha}]\ket{\alpha,n}$  where $\alpha$ denotes the state of the qubit, which can be either up $(\alpha=1)$ or down $(\alpha=-1)$, and $n$ denotes the Fock state of the cavity. Here $S[r_{\alpha}]={\rm exp}[r_\alpha(\hat{c}^{\dagger 2}-\hat{c}^{2})/2]$ is the squeezing operator with the squeeze parameter $r_{\alpha}=-\frac{1}{4}{\rm ln}(1+\alpha g^2)$. The respective eigenenergies are $E_{\alpha,n}=n \epsilon_{\alpha}+\frac{\epsilon_{\alpha}-\omega}{2}+\alpha \frac{\Omega}{2}$ with $\epsilon_{\alpha}=\omega \sqrt{1+\alpha g^2}$.

In the eigenbasis of the Rabi Hamiltonian, the matrix elements of the operator $\hat{A}(t)$ are expressed as
\begin{equation} \label{eq:matrix_elements}
\begin{split}
&\bra{E_{\alpha,n}} \hat{A}(t) \ket{E_{\beta,m}}=\\
&\delta_{\alpha,\beta}e^{-i \nu_0 t} ({\rm cosh} r_{\alpha}\sqrt{m} \delta_{n,m-1}+{\rm sinh}r_{\alpha}\sqrt{m+1} \delta_{n,m+1})\\
+&\delta_{\alpha,\beta}e^{i \nu_0 t} ({\rm sinh}r_{\alpha}\sqrt{m} \delta_{n,m-1}+{\rm cosh}r_{\alpha}\sqrt{m+1} \delta_{n,m+1})
\end{split}
\end{equation}
where $\delta_{i,j}$ is the Kronecker delta function. As a result, the interaction-picture operator $\tilde{A}(t)=\sum \limits_{n,m,\alpha,\beta} e^{i (E_{\alpha,n}-E_{\beta,m})t} \bra{E_{\alpha,n}} \hat{A}(t) \ket{E_{\beta,m}} \ket{E_{\alpha,n}} \bra{E_{\beta,m}}$ takes the form
\begin{equation} \label{eq:interaction_term}
\begin{split}
\tilde{A}(t)=&\sum_{\alpha} \ket{\alpha}\bra{\alpha} S[r_{\alpha}] \hat{c} S^{\dagger}[r_{\alpha}] {\rm cosh}r_{\alpha} e^{-i (\nu_0+\epsilon_{\alpha}) t}\\
+&\sum_{\alpha} \ket{\alpha}\bra{\alpha} S[r_{\alpha}] \hat{c} S^{\dagger}[r_{\alpha}] {\rm sinh}r_{\alpha} e^{i (\nu_0-\epsilon_{\alpha}) t}\\
+&\sum_{\alpha} \ket{\alpha}\bra{\alpha} S[r_{\alpha}] \hat{c}^{\dagger} S^{\dagger}[r_{\alpha}] {\rm sinh}r_{\alpha} e^{-i (\nu_0-\epsilon_{\alpha}) t}\\
+&\sum_{\alpha} \ket{\alpha}\bra{\alpha} S[r_{\alpha}] \hat{c}^{\dagger} S^{\dagger}[r_{\alpha}] {\rm cosh}r_{\alpha} e^{i (\nu_0+\epsilon_{\alpha}) t}.\\
\end{split}
\end{equation} 
The last step of our microscopic derivation is to insert Eq.~\eqref{eq:interaction_term} and the bath correlation function $J(s)$ into Eq.~(\ref{eq:me_interact}) and perform the secular approximation. As concrete examples, in the following we consider two different environments.

\subsection{Case 1: bosonic bath at zero temperature as the environment}\label{sec:thermal_bath}
In the first case we consider that the cavity-qubit system interacts with a bosonic bath at zero temperature characterized by $\hat{H}_E= \sum \limits_ k \omega_k \hat{b}_k^{\dagger} \hat{b}_k$ and $\hat{B}= \sum \limits _k \gamma_k (\hat{b}_k+\hat{b}_k^{\dagger})$, where $\omega_k$ is the frequency of the bosonic modes and $\gamma_k$ is the coupling strength between the system and the environment. In this case the spectral density can be derived to be $\Gamma(\omega)=f(\omega)|\gamma(\omega)|^2$, with $f(\omega)$ being the density of states, and satisfies 
\begin{equation} \label{eq:fourier}
\Gamma(\omega)
\begin{cases}
>0, & \omega \geq 0, \\
= 0,  &\omega < 0.
\end{cases}
\end{equation}
We further assume $\nu_0 \gg \epsilon_\alpha$ ($\alpha= \pm 1$) and that $\Gamma(\omega)$ is sufficiently flat in the narrow frequency window $[\nu_0-\epsilon_\alpha,\nu_0+ \epsilon_\alpha]$ in accordance with our Markov assumption. Under these assumptions, only the first and third terms of Eq.~(\ref{eq:interaction_term}) gives a nonzero contribution to the first line of Eq.~(\ref{eq:me_interact}), while the second and fourth terms of Eq.~(\ref{eq:interaction_term}) gives a nonzero contribution to the second line of Eq.~(\ref{eq:me_interact}). Furthermore, in the spirit of the partial-secular approximation~\cite{doi:10.1063/1.4907370,PhysRevA.100.012107,PhysRevA.101.042108}, we can neglect all the fast rotating terms at frequency approximately $ \pm 2 \nu_0 $ and keep only the terms that oscillate with frequencies $0$, $\pm 2 \epsilon_{\pm 1}$, $\pm(\epsilon_1-\epsilon_{-1})$, $\pm (\epsilon_1+\epsilon_{-1})$. As $\Gamma(\omega)$ is sufficiently flat around $\nu_0$, we have $\Gamma(\nu_0 \pm \epsilon_{1})=\Gamma(\nu_0 \pm \epsilon_{-1}) \equiv\kappa$. 

Inserting Eq. (\ref{eq:interaction_term}) into Eq. (\ref{eq:me_interact}) and moving back to the Schr\"{o}dinger picture, the resulting terms can be conveniently regrouped. Importantly, the resulting jump operator is exactly the annihilation operator $\hat{c}$ of the cavity mode. As an example, the integral $\int_0^\infty dsJ(s)\tilde{A}(t-s)\tilde{\rho}(t)\tilde{A}(t)$ in Eq. (\ref{eq:me_interact}) leads to the following term in the Schr\"odinger picture,
\begin{equation} 
\begin{split}
&\int_0^\infty dsJ(s)\tilde{A}(t-s)\tilde{\rho}(t)\tilde{A}(t) \rightarrow  \\
&\frac{\kappa}{2} \left(\sum \limits_ {\alpha} \ket{\alpha}\bra{\alpha} S[r_{\alpha}] \hat{c} S^{\dagger}[r_{\alpha}] {\rm cosh}r_{\alpha} +\right. \\
& \left. \sum \limits_{\alpha} \ket{\alpha}\bra{\alpha} S[r_{\alpha}] \hat{c}^{\dagger} S^{\dagger}[r_{\alpha}] {\rm sinh}r_{\alpha} \right) \rho  \\
& \left(\sum \limits_{\beta} \ket{\beta}\bra{\beta} S[r_{\beta}] \hat{c} S^{\dagger}[r_{\beta}] {\rm sinh}r_{\beta} +\right. \\
& \left. \sum \limits_{\beta} \ket{\beta}\bra{\beta} S[r_{\beta}] \hat{c}^{\dagger} S^{\dagger}[r_{\beta}] {\rm cosh}r_{\beta} \right)=\frac{\kappa}{2} \hat{c} \rho \hat{c}^\dag,
\end{split}
\end{equation} 
Here we have used the facts that $\sum \limits_ {\alpha} \ket{\alpha}\bra{\alpha}=\hat{I} $ with $\hat{I}$ being the identity operator in the spin Hilbert space, and that ${\rm cosh}r_\alpha\hat{c} + {\rm sinh}r_\alpha\hat{c}^{\dag} =S^{\dagger}[r_\alpha] \hat{c} S[r_\alpha] $. As a result, by including all of the terms of Eq. (\ref{eq:me_interact}) we arrive at the LME in the Schr\"{o}dinger picture
\begin{equation}
\dot{\rho}= -i [\hat{H},\rho]+\kappa\left(\hat{c} \rho \hat{c}^\dag -\frac{1}{2}\{\hat{c}^\dag\hat{c}, \rho\} \right),
\end{equation}
which recovers Eq.~\eqref{eq:LME_Rabi} of the main text.

\subsection{Case 2: ancilla ion as the environment} \label{sec:ancilla_ion}
As a second case, we examine the dynamics of the proposed experimental setup in Sec (\ref{sec:Experimental_implementation}), where a second ancilla ion is confined in the same trap as the first ion, driven by a laser with Rabi frequency $\Omega_{\rm r}$ and detuning $\Delta_{\rm e}$ on the cooling transition $\ket{g}\to\ket{e}$ with spontaneous emission rate $\Gamma_{\rm e}$. In the Lamb-Dicke limit, this experimental setup leads to an interaction Hamiltonian between the cavity mode (realized as phonons) and the ancilla ion of the form $\hat{H}_{ \rm int}=e^{i \nu_0t \hat{c}^{\dagger} \hat{c}}(\ha+\hadag) e^{-i \nu_0t \hat{c}^{\dagger} \hat{c}} \otimes (\ket{e} \bra{g}+ \ket{g}\bra{e})$. Under the weak coupling $\eta_{\rm LD}\Omega_{\rm r} \ll \Gamma_{\rm e}$ assumption, the environmental noise-power spectrum has the simple expression \cite{PhysRevA.48.2169}
\begin{equation} \label{eq:spectrum_ancilla}
\Gamma(\omega)\approx \eta_{\rm LD}^2 \Omega_{r}^2 \frac{\Gamma_{e}}{\Gamma_{ e}^2+(\omega+\Delta_{ e})^2}.
\end{equation}
Typical experiment satisfies the resolved sideband condition $\Gamma_e \ll \omega_{\rm ph}\sim \nu_0$ such that the spectrum density (\ref{eq:spectrum_ancilla}) is well localized around its center $\omega=- \Delta_e$. Therefore, if we tune the laser to be on resonance with the red phonon sideband, $\Delta_{e}=- \omega_{\rm ph}$, we have a situation similar to the first case where only the first and third terms of Eq.(\ref{eq:interaction_term}) gives a nonzero contribution to the first line of Eq. (\ref{eq:me_interact}), while the second and fourth terms of Eq.(\ref{eq:interaction_term}) gives a nonzero contribution to the second line of Eq. (\ref{eq:me_interact}). Likewise, we can perform the semi secular approximation by keeping the terms rotating with frequencies $0$, $\pm 2 \epsilon_{\pm 1}$, $\pm(\epsilon_1-\epsilon_{-1})$, $\pm (\epsilon_1+\epsilon_{-1})$ and neglecting the terms rotating at frequency approximately $ \pm 2 \nu_0 $. In typical experiment $\Gamma_e \gg \epsilon_{\alpha}$, such that $\Gamma(\omega)$ is sufficiently flat and, correspondingly, we can approximate $\Gamma(\nu_0 \pm \epsilon_{1})=\Gamma(\nu_0 \pm \epsilon_{-1}) \equiv \kappa$. Regrouping the relevant terms we obtain Eq.~(\ref{eq:LME_Rabi}) of the main text.

\subsection{LME for finite effective system size $\eta$}
Above we have derived the LME (\ref{eq:LME_Rabi}) based on the microscopical Hamiltonians of our proposed trapped-ion setup in the thermodynamic limit $\eta \rightarrow \infty$. Here we discuss the leading order corrections to Eq.~(\ref{eq:LME_Rabi}) at finite and large $\eta \gg 1$. This can be done by carrying out the same derivation as before, using the eigenstates of the Rabi Hamiltonian Eq.~(\ref{eq: Hamiltonian}) at finite $\eta$. For example, for $g<1$, such eigenstates are~\cite{PhysRevLett.115.180404}
\begin{equation}
\ket{\tilde{E}_{\alpha,n}}=\hat{U} \ket{E_{\alpha,n}}+\mathcal{O} \left(\frac{1}{\eta}\right),
\end{equation}
where $\hat{U}={\rm exp} \left[ \frac{g}{2 \sqrt{\eta}}(\hat{c}+\hat{c}^{\dagger})(\hat{\sigma}_{+}-\hat{\sigma}_-) \right]$ and $\ket{E_{\alpha,n}}$ are the eigenstates of Eq.~(\ref{eq: Hamiltonian}) when $\eta\to\infty$ (see Appendix \ref{sec:expansion_eigenstates}). Using these eigenstates, the LME for finite $\eta$ can be derived to be
\begin{align}
\label{eq:l_corr}
\dot{\rho}=\tilde{\mathcal{L}} \rho &=-i [\hat{H},\rho]+{\cal D}[\hat{U} \hat{c} \hat{U}^{\dagger}] \rho+ \mathcal{O}\left(\frac{1}{\eta}\right)\nonumber \\
& \equiv \mathcal{L} \rho + \mathcal{L}_{\rm corr} \rho,
\end{align}
where we have defined ${\cal D}[\hat{F}] \rho= \hat{F} \rho \hat{F}^{\dagger}-\{ \hat{F}^{\dagger} \hat{F}, \rho \}/2$ and $\mathcal{L}$ is given by the right-hand side of Eq.~\eqref{eq:LME_Rabi}]. The same equation can be derived for $g\geq1$.

Dissipative criticality is defined via gap closing of $\cal L$ at the CP $g=g_{\rm CP}$; cf. Sec.~\ref{sec:Quantum_Rabi}. We have derived analytically the scaling of the perturbation term $\mathcal{L}_{\rm corr}$ in Eq.~\eqref{eq:l_corr} with respective to $\eta$ at the CP $g=g_{\rm CP}$, given by $\mathcal{L}_{\rm corr} \sim \frac{g}{2 \sqrt{\eta}} (\ha+\hadag) (\hat{\sigma}_{+}-\hat{\sigma}_-) \mathcal{L}\sim \eta^{-1/4} \mathcal{L}$ (we note that $\langle \ha+\hadag \rangle \sim \eta^{-1/4}$ and $\langle \hat{\sigma}_{\pm} \rangle \sim \eta^0 $), which demonstrate that $\mathcal{L}_{\rm corr}$ has a negative scaling dimension. Therefore, $\mathcal{L}_{\rm corr}$ serves as an irrelevant perturbation to the criticality, which barely shifts the location of the CP, and leaves all the critical properties---including the critical scalings of the global QFI derived in the main text---unchanged.
\bibliography{OpenRabiModel_bib}
\end{document}